\documentclass[twocolumn]{article}    
\usepackage[left=1.5cm, right = 1.5cm, top = 2.5cm, bottom = 2.5cm]{geometry}

\usepackage[utf8x]{inputenc}

\usepackage{graphicx}
\usepackage{latexsym,amssymb,amsfonts,graphicx,xcolor, amsmath,amsthm}

\usepackage{url}\RequirePackage[colorlinks,citecolor=blue, linkcolor=blue,urlcolor = blue]{hyperref}

\usepackage{algorithm}
\usepackage[noend]{algpseudocode}
\usepackage{rotating}
\usepackage{pdflscape}

\usepackage{authblk} 

\newtheorem{theorem}{Theorem}[]
\newtheorem*{theorem*}{Theorem}

\newtheorem*{claim*}{Claim}
\theoremstyle{definition}
\newtheorem{definition}[theorem]{Definition}
\newtheorem*{definition*}{Definition}
\theoremstyle{remark}

\newtheorem*{example*}{Example}

\newcommand{\eps}{\ensuremath\epsilon}
\newcommand{\kap}{\ensuremath\kappa}
\newcommand{\ka}{\ensuremath\kappa_1}
\newcommand{\kb}{\ensuremath\kappa_2}
\newcommand{\R}{\ensuremath\mathbb{R}}

\newcommand{\W}{\ensuremath\Omega}

\newcommand{\bord}{\ensuremath \partial}
\newcommand{\hauss}{\ensuremath \mathcal{H}}
\newcommand{\inter}{\ensuremath \cap}
\newcommand{\union}{\ensuremath \cup}

\renewcommand{\S}{\ensuremath\mathcal{S}}

\newcommand{\n}{\ensuremath\mathbf{n}}
\newcommand{\abf}{\ensuremath\mathbf{a}}
\newcommand{\bbf}{\ensuremath\mathbf{b}}

\newcommand{\Ebf}{\ensuremath\mathbf{E}}
\newcommand{\Fbf}{\ensuremath\mathbf{F}}

\newcommand{\Lap}{\ensuremath\Delta}
\newcommand{\Div}{\ensuremath\nabla \cdot}
\newcommand{\grad}{\ensuremath\nabla}
\newcommand{\ngd}{\ensuremath |\grad u|}

\newcommand{\Hess}{\ensuremath ~\mathrm{Hess}~}

\newcommand{\Beps}{\ensuremath \mathcal{B}^\eps_u}

\newcommand{\Bueps}{\ensuremath \mathcal{B}^\eps_{u_\eps}}
\newcommand{\Heps}{\ensuremath \mathcal{H}^\eps_u}
\newcommand{\Keps}{\ensuremath \mathcal{K}^\eps_u}
\newcommand{\Meps}{\ensuremath \mathcal{M}^\eps_u}
\newcommand{\Mueps}{\ensuremath \mathcal{M}^\eps_{u_\eps}}

\newcommand{\kepsa}{\ensuremath \kap_{1,u}^\eps}
\newcommand{\kepsb}{\ensuremath \kap_{2,u}^\eps}

\newcommand{\Hel}{\ensuremath \mathbf{E}_\text{H}}
\newcommand{\Will}{\ensuremath \mathbf{E}_\text{W}}
\newcommand{\Eeps}{\ensuremath\mathcal{E}_\eps}
\newcommand{\Feps}{\ensuremath\mathcal{F}_\eps}

\newcommand{\Tra}[1]{\ensuremath{\mathrm{Tr} #1} }
\newcommand{\Nor}[1]{\ensuremath{ \| #1 \| } }

\begin{document}

\title{\textbf{Generation of tubular and membranous shape textures\\ with curvature functionals}}

\author[1, 2,*]{Anna Song}
\affil[1]{\small Department of Mathematics, Imperial College London, UK}
\affil[2]{\small Haematopoietic Stem Cell Laboratory, The Francis Crick Institute, UK }
\affil[*]{\small Corresponding e-mail: a.song19@imperial.ac.uk}


\maketitle

\begin{abstract}
Tubular and membranous shapes display a wide range of morphologies that are difficult to analyze within a common framework.
By generalizing the classical Helfrich energy of biomembranes, we model them as solutions to a curvature optimization problem in which the principal curvatures may play asymmetric roles.
We then give a novel phase-field formulation to approximate this geometric problem, and study its Gamma-limsup convergence. 
This results in an efficient GPU algorithm that we validate on well-known minimizers of the Willmore energy; the software for the implementation of our algorithm is freely available online.
Exploring the space of parameters reveals that this comprehensive framework leads to a wide continuum of shape textures.
This first step towards a unifying theory will have several implications, in biology for quantifying tubular shapes or designing bio-mimetic scaffolds, but also in computer graphics or architecture.
\end{abstract}


\section{Introduction}
\label{sec:intro}


Tubular and membranous shape textures are widely present in biology. They display a large variety of morphologies in terms of geometry and topology, which are important to analyze since they reflect the state of a biological system. For instance, the bone marrow capillaries are highly branching and merging vessels \cite{sivaraj_blood_2016,ramasamy_structure_2017}, whose organization is subject to drastic remodeling in acute myeloid leukaemia \cite{passaro_increased_2017,duarte_inhibition_2018}. In cells, the endoplasmic reticulum, where proteins are synthesized, consists of an interconnected network \cite{schwarz_endoplasmic_2016} that undergoes sheets-to-tubules topological transformations \cite{puhka_progressive_2012}. Furthermore, trabecular bone is a combination of rods and platelets \cite{muller_hierarchical_2009,salmon_structure_2015} that are optimally restructured under mechanical stress \cite{ryan_unique_2012,acquaah_early_2015,salmon_non-linear_2015} or pathological conditions \cite{parfitt_trabecular_1987,tamada_three-dimensional_2005,frost_modeling-based_2020}.

\begin{figure*}
	\centering
	\includegraphics[trim = 0 0 0 0cm, clip, width=1\textwidth]{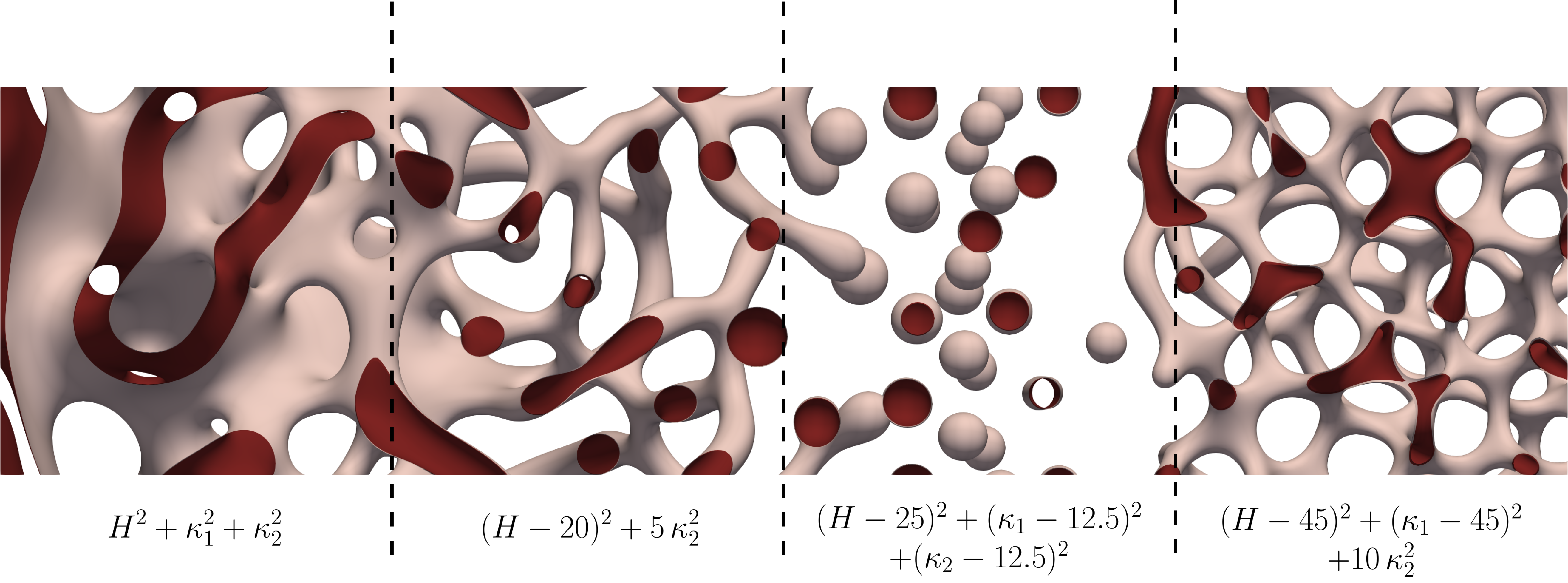} 
	\caption{\small\textbf{A continuum of shape textures generated by curvatubes,} after optimizing a curvature-based polynomial energy of the surface $\int_\S p(\ka,\kb) ~dA$, with a volume constraint. The polynomial coefficients vary linearly in space, by interpolating four values that define different shape textures, at the vertical median of the four squares. In the formulas, $\ka$ and $\kb$ are the principal curvatures of the suface, while $H = \ka + \kb$ and $K = \ka \kb$ are the mean and the Gaussian curvatures. More details are given at the end of Section \ref{sec:discu_extensions}.}
	\label{fig:spatialized}
\end{figure*}

However, due to their disparity and complexity, tubular and membranous structures are difficult to describe within a \textit{unifying framework} that captures both their rich morphological diversity as well as their continuous variations.
We approach this question by building a \textit{generation model} that creates \textit{shape textures} from noise, similarly to texture synthesis in images \cite{julesz_visual_1962,portilla_parametric_2000,landy_visual_2004}.
We model tubules and membranes as optimizers under constant volume of a \textit{curvature functional}
\begin{equation} \label{eq:generalk1k2}
\Fbf(\S) = \int_\S p(\ka, \kb) ~ dA,
\end{equation}
where $p$ is a second-degree polynomial of the principal curvatures $\ka$ and $\kb$ of the surface $\S$. As our main contribution, we provide a novel phase-field formulation $\Feps$ to approximate the original geometric problem $\Fbf$, and show that the $\Gamma$-limsup holds, a notion coming from the $\Gamma$-convergence framework \cite{de_giorgi_remarks_1991,alberti_variational_2000,braides_gamma-convergence_2002}. The optimization problem then translates into the mass-preserving $H^{-1}$ gradient flow \cite{evans_partial_2010,cowan_cahn-hilliard_2005}
\begin{equation} \label{eq:Hminus_flow}
\dot{u} = \Lap \frac{\partial \Feps}{\partial u}.
\end{equation}
Combining the stochastic optimizer Adam \cite{kingma_adam_2015,loshchilov_decoupled_2019} to the automatic differentiation provided by PyTorch \cite{paszke_pytorch_2019} results in an efficient and flexible GPU implementation, \verb|curvatubes|. It successfully leads to a wide continuum of shape textures (see Figure \ref{fig:spatialized}), which constitutes a first step towards a unifying theory.

\paragraph{Related work.}
The variational formulation \eqref{eq:generalk1k2} generalizes two classical functionals, the \textit{Willmore energy} studied in differential geometry \cite{simon_existence_1993,willmore_riemannian_1996,bauer_existence_2003,marques_min-max_2014,keller_embedded_2014,bretin_phase-field_2015,toda_willmore_2018}, as well as the \textit{Helfrich energy} used to model biomembranes \cite{canham_minimum_1970,helfrich_elastic_1973,deuling_red_1976,du_phase_2004,seguin_microphysical_2014,campelo_helfrich_2014}. In contrast to these models, the polynomial $p$ is here not required to be symmetric in the principal curvatures, which allows the generation of tubules. Other generalizations have been proposed in \cite{tu_geometric_2004,gruber_variation_2019} and \cite{dogan_first_2012,chicco-ruiz_shape_2018}.

Curvature functionals are often used as image prior models in imaging, due to their ability to interpolate. Mumford \cite{mumford_elastica_1994} considered Euler's elastica functional as a prior curve model in computer vision, and this was subsequently applied to digital inpainting \cite{masnou_level_1998,bertalmio_image_2000,shen_eulers_2003}. Similar ideas were then used for 3D volume reconstruction from 2D slices \cite{li_three-dimensional_2015,bretin_volume_2017,kim_three-dimensional_2019}. Our work is also related to Poisson reconstruction methods \cite{kazhdan_poisson_2006,kazhdan_screened_2013} that implicitly reconstruct a surface from noisy oriented points.

However, as in \cite{du_phase_2004,du_retrieving_2005,du_diffuse_2007}, we use the diffuse setting to model shapes.
By construction, our phase-field formulation $\Feps$ extends the standard approximation of the Willmore energy (see \cite{bretin_phase-field_2015} and references therein) as well as the one proposed by Bellettini \& Mugnai in \cite{bellettini_approximation_2010} for the Helfrich energy.
In terms of $\Gamma$-convergence, we also provide an extension of the $\Gamma$-limsup result in \cite{bellettini_approximation_2010}, while keeping the $\Gamma$-liminf as an open question (except in the cases previously covered) that we do not aim to solve here.

Finally, our attempt in building a unifying framework is connected to the Functionalized Cahn--Hilliard (FCH) model proposed in \cite{gavish_variational_2012,dai_geometric_2013,christlieb_high_2014,kraitzman_overview_2015,kraitzman_pearling_2017,christlieb_competition_2019}. The FCH energy describes how amphiphilic molecules self-assemble into complex network morphologies that feature spheres, tubules, sheets, and mixtures of them. Loosely speaking, such shape textures result from a compromise between minimizing a bending energy while rewarding an increase in interfacial area. 
Yet, we choose to treat the principal curvatures independently, which is not the case in their model.

\paragraph{Outline.}
In Section \ref{sec:background}, we introduce the Willmore and the Helfrich energies, as well as their classical phase-field approximations in light of the $\Gamma$-convergence framework. In section \ref{sec:model}, we build the phase-field functional and justify its construction with heuristic arguments and in terms of $\Gamma$-limsup.
Section \ref{sec:simulations} describes the computational framework and displays a large number of simulations.
Finally, we discuss in Section \ref{sec:discussion} the implications of a unifying theory on applied fields and present future extensions to this work.


\section{Background}
\label{sec:background}

We begin by introducing the Willmore and the Helfrich energies in more detail, then sketch the notion of $\Gamma$-convergence before moving on with their classical phase-field approximations.

But first, let us describe curvatures in simple terms. \textit{Curvature} measures how much a line or a surface is locally deviating from a straight line or a flat plane. For a line contained in a plane, the curvature $\kap$ at a point is the inverse $\frac{1}{r}$ of the radius of the osculating circle, i.e., the tangent circle that approaches the curves most tightly at this point. Surfaces are characterized at each point by two \textit{principal curvatures} $\kap_1$ and $\kap_2$, that correspond to the maximal and minimal curvatures of the lines resulting from the perpendicular intersection of the surface with a plane. The maximally and minimally curved lines are directed by two perpendicular \textit{principal directions}, tangent to the surface. For instance, the principal curvatures are zero on a plane; on a sphere of radius $r$, they are all equal to $(\frac{1}{r}, \frac{1}{r})$; on a cylinder of base radius $R$, they are all equal to $(\frac{1}{R}, 0)$.

\subsection{The Willmore, Helfrich, and generalized curvature functionals} 
In the 1970's, Canham \cite{canham_minimum_1970}, and subsequently Helfrich \cite{helfrich_elastic_1973}, proposed to model the surface of a biomembrane as a minimizer of a curvature bending energy, or \textit{Helfrich energy}
\begin{equation}\label{eq:Helfrich}
\Hel(\S) = \int_{\S} \left( \frac{\chi_b}{2}(H - H_0)^2 + \chi_G K \right) \,dA.
\end{equation}
In this expression, $\S$ is a smooth compact orientable surface in $\R^3$ 
whose principal curvatures are denoted $\ka$ and $\kb$, with the convention $\ka \geq \kb$. The mean curvature $H = \kap_1 + \kap_2$ and the Gaussian curvature $K = \kap_1 \kap_2$ are respectively the sum and the product of the principal curvatures. The signs of $H$, $\ka$, and $\kb$ depend on the orientation of $\S$, while that of $K$ does not. Throughout this work, we use the convention that $H$ should be positive on convex objects like spheres. Let us remark that there is a closed relationship between $(\ka,\kb)$ and $(H,K)$, provided by the bijection $\phi(x,y) = (x+y, \, x \, y)$ from $\{(x,y) \in \R^2 ~|~ y \leq x \}$ onto $\{(x,y) \in \R^2 ~|~ y \leq \frac{1}{4} \, x^2 \}$, 
\begin{equation}
\label{eq:HK_kakb}
\begin{cases}
H = \ka + \kb \\
K = \ka \kb \\
\ka \geq \kb
\end{cases}
\Leftrightarrow \quad
\begin{cases}
\ka = ( H + \sqrt{H^2 - 4 K} ) / 2 \\
\kb = ( H - \sqrt{H^2 - 4 K} ) / 2 \\
\end{cases}
\end{equation}
and that $H^2 - 4 K = (\ka - \kb)^2$.

In \eqref{eq:Helfrich}, the coefficients $\chi_b > 0$ and $\chi_G$ are the bending and Gaussian rigidities. The parameter $H_0$, or spontaneous curvature, models the asymmetry of the two layers composing the bilipidic membrane \cite{seifert_configurations_1997,dobereiner_spontaneous_1999}. 
The \textit{Willmore energy}, defined as
\begin{equation}\label{eq:Willmore}
\Will(\S) = \int_{\S} H^2 ~dA,
\end{equation}
is then a special case of the Helfrich energy, with $\chi_G = 0$, $\chi_b = 2$, and $H_0 = 0$.

In both of the classical functionals \eqref{eq:Helfrich} and \eqref{eq:Willmore}, the integrand is a polynomial $p(\ka,\kb)$ symmetric in the principal curvatures, i.e., $p(\ka,\kb) = p(\kb,\ka)$, since it can be reformulated as a polynomial of their sum $H$ and product $K$.
However, one may want to construct a general curvature functional where $p$ is a smooth function with no symmetry constraint. A similar\footnote{Yet, please note that $p$ being smooth in $(\ka,\kb)$ is not equivalent to $q$ being smooth in $(H,K)$, where $q = p \circ \phi^{-1}$. It can be checked by taking $q(H,K) = 2 \ka = H + \sqrt{H^2 - 4 K}$, which is not differentiable at points where $H^2 = 4 K$, i.e., when $\ka = \kb$. On a surface, this happens at umbilical points, e.g., everywhere on spheres.} form is given in \cite{tu_geometric_2004,toda_willmore_2018}, for a smooth function $q(H,K)$. A generalization to functions that depend on the position and the normal to the surface can be found in \cite{dogan_first_2012,chicco-ruiz_shape_2018}. 

We will restrict ourselves to $p$ which are polynomials of degree $2$, as the framework is then rich enough to generate complex shape textures. We are thus interested in the curvature functional
\begin{align}
\label{eq:F}
\begin{split}
\Fbf(\S) &= \int_{\S} \big(a_{2,0} ~\kap_1^2 + a_{1,1} ~\kap_1 \kap_2 + a_{0,2} ~\kap_2^2  \\
& \hskip1cm + a_{1,0} ~\ka + a_{0,1} ~\kb + a_{0,0} \big) ~dA \\
&= \int_\S \left( \sum \limits_{|\alpha| \leq 2} a_\alpha (\ka,\kb)^\alpha \right) ~dA,
\end{split}
\end{align}
where we use multi-index notation.

\subsection{Phase-fields and $\Gamma$-convergence}
\label{sec:gamma_background}

Numerically, critical points of the curvature energies \eqref{eq:Helfrich}, \eqref{eq:Willmore}, and \eqref{eq:F} can be searched for by following a gradient flow. Before implementing an algorithm, it is however preferable to convert these sharp-interface functionals defined for 2D surfaces, to diffuse approximations defined for scalar fields in a 3D volume. This way, surfaces are implicitly represented as level sets of the volumetric scalar field, which allows us to address topological changes encountered in the flow seamlessly; whereas in explicit methods, surfaces are tracked as a mesh that needs dynamic remeshing to avoid entanglement through topological transitions \cite{osher_level_2002}. The gain is considerable in view of the high topological complexity of the targeted shape textures.

This leads us to consider a \textit{phase-field} approximating the original geometric problem, for instance as in \cite{du_phase_2004,du_retrieving_2005,du_diffuse_2007} for the modeling of biomembranes. Phase-fields have been extensively used to model phase separation in binary mixtures, beginning with the Cahn--Hilliard (or Ginzburg--Landau)
energy \cite{cahn_free_1958,elliott_cahn-hilliard_1989}, that was subsequently reused in several other contexts \cite{kim_basic_2016,miranville_cahnhilliard_2017}.  
These functions typically take values close to $1$ and $-1$ inside and outside a region, with a smooth transition between the two phases at the interface. A 
parameter $\eps > 0$ represents the thickness of phase transition at the interface. \newline

The quality of a diffuse approximation with phase-fields is typically studied in the $\Gamma$-convergence framework \cite{de_giorgi_remarks_1991,alberti_variational_2000,braides_gamma-convergence_2002}. $\Gamma$-convergence expresses the convergence of minimization problems, so that, rather than solving a limit problem, we solve a sequence of approaching problems (or the reverse). The complete convergence consists in a $\Gamma$-limsup, which relies on a constructive proof, and a $\Gamma$-liminf, generally more difficult to prove.
\begin{definition}[$\Gamma$-convergence]
	Given $X$ a metric space, let $F$ and $F_\eps$ be functions from $X$ to $[-\infty,+\infty]$, where the $F_\eps$ are defined for $\eps > 0$.
	We say that the sequence $(F_\eps)_{\eps > 0}$ $\Gamma$-converges to $F$ at a point $u \in X$ as $\eps \to 0$, and write $$\left[\Gamma(X) - \lim\limits_{\eps \to 0} F_\eps\right] (u) = F(u),$$ if the following two bounds hold:
	
	\noindent \textbf{[$\Gamma$-liminf]} For every sequence $(u_\eps)$ such that $u_\eps \to u$ in $X$,
	$$ \liminf\limits_{\eps \to 0} F_\eps(u_\eps) \geq F(u). $$
	
	\noindent \textbf{[$\Gamma$-limsup]} There exists a sequence $(u_\eps)$, called \textit{recovery sequence}, such that $u_\eps \to u$ in $X$ and
	$$\lim\limits_{\eps \to 0} F_\eps(u_\eps) = F(u).$$
\end{definition}
This punctual definition can be extended to a convergence taking place on the whole space. 
$\Gamma$-convergence is especially interesting due to the following fundamental result.
\begin{theorem}
	Let X be a metric space, and let $F = \Gamma(X)-\lim\limits_{\eps \to 0} F_\eps$. Suppose that the sequence $F_\eps : X \to [-\infty, +\infty]$ is equi-coercive, i.e., for all $t \in \R$ there exists a compact set $K_t \subset X$ such that $\{ F_\eps \leq t \} \subset K_t$. Then $F$ admits a minimum and
	$$\min\limits_X F = \lim\limits_{\eps \to 0} \inf\limits_X F_\eps.$$
	Futhermore, if $u_\eps$ minimizes $F_\eps$ over $X$, then every cluster point of $(u_\eps)$ minimizes $F$ over $X$. 
\end{theorem}
This ensures not only the convergence of the minimal values, but also of the minimizers themselves.

\subsection{Phase-field approximations of the area, the Willmore, and the Helfrich functionals}
\label{sec:classical_phase_fields}
In this paragraph, we present three classical diffuse approximations that are important to our development. We fix some mathematical notations beforehand.
\paragraph{Notations.}
Let $\W$ denote an open \textit{bounded} connected set in $\R^3$ with smooth boundary. The usual Sobolev spaces are denoted by $W^{k,p}(\W) = \{ u \in L^p(\W) ~|~ D^\alpha u \in L^p(\W), \forall |\alpha| \leq k \}$ and are the sets of functions u in $L^p(\W)$ whose mixed partial derivatives $D^\alpha u$ exist in the weak sense and are in $L^p(\W)$, up to $|\alpha| \leq k$. By $BV(\W,\{-1,1\})$, we denote the set of functions $u : \W \to \{-1,1\}$ such that $u$ is of bounded variation, i.e., $u \in L^1(\W)$ and $\int_\W |Du| < +\infty$. For a set $E \subset \W$, $\chi_E$ designates the characteristic function of $E$.

Let $\mathbf{e}$ be a fixed unit-norm vector in $\R^3$.
We consider a symmetric double-well function $W(s) = \frac{1}{4} (1-s^2)^2$ that cancels on $-1$ and $+1$. Note that its derivative is $W'(s) = s^3 - s$. Let $\sigma = \int_{-1}^{1} \sqrt{2 W(s)} ~ds = \frac{4}{3 \sqrt{2}}$ denote a constant that only depends on the double-well.

For a function $u$ twice (weakly) differentiable, we define the normal vector field
\begin{equation*}
\n_u = \left\{ \begin{array}{cc}
\frac{\grad u}{|\grad u|} & \text{on the set } \{ \grad u \neq 0 \} \\
\mathbf{e} & \text{elsewhere},
\end{array} \right.
\end{equation*}
that has unit norm, and is orthogonal to the level sets of $u$. We introduce the matrix field
\begin{equation}
\label{eq:Meps}
\Meps = - \eps \Hess u + \frac{W'(u)}{\eps} ~ \n_u \otimes \n_u,
\end{equation}
whose trace is equal to 
\begin{equation*}
\Tra{\Meps} = -\eps \Lap u + \frac{W'(u)}{\eps}.
\end{equation*}

\paragraph{The classical approximations.}

The Cahn--Hilliard phase-field is known to approximate the \textit{area} (or perimeter) \textit{functional}, which measures the total area of surfaces in the 3D space. More precisely, let us introduce 
\begin{equation}
\label{eq:CH_eps}
\mathcal{A}_\eps(u) = 
\begin{cases}
\displaystyle \int_\W \left(\frac{\eps}{2} \ngd^2 + \frac{W(u)}{\eps} \right) ~dx ~~\text{if } u \in W^{1,2}(\W) \\ 
+ \infty ~~\text{otherwise in } L^1(\W)
\end{cases},
\end{equation}
and the area functional
\begin{equation}
\Ebf_A(u) = 
\begin{cases}
\frac{1}{2} |Du|(\W) ~~\text{if } u \in \mathrm{BV}(\W, \{ -1,1 \}) \\
+ \infty ~~\text{otherwise in } L^1(\W)
\end{cases}.
\end{equation}
Following a conjecture of De Giorgi, Modica and Mortola \cite{modica_esempio_1977} proved the $\Gamma$-convergence
\begin{equation}
\left[\Gamma(L^1(\W)) - \lim\limits_{\eps \to 0} \mathcal{A}_\eps\right]= \sigma\,\Ebf_A.
\end{equation}
This means that, if $E \subset \W$ is such that $\S = \bord E \inter \W$ is smooth and of finite area, and setting $u = 2 \chi_E - 1 \in BV(\W,\{-1,1\})$, then the $\Gamma$-limsup provides a sequence of functions $(u_\eps) \in W^{1,2}(\W)$ such that $u_\eps \to u$ in $L^1(\W)$ and $\mathcal{A}_\eps(u_\eps) \to \sigma \, \text{area}(\S)$, i.e., their diffuse areas converge to the area of $\S$ up to a factor $\sigma$.\newline

Subsequently, several authors \cite{bellettini_approssimazione_1993,tonegawa_phase_2002,bellettini_approximation_2005,moser_higher_2005,roger_modified_2006,nagase_singular_2007} studied diffuse approximations of the Willmore energy \eqref{eq:Willmore}. Bellettini and Paolini \cite{bellettini_approssimazione_1993} introduced the phase-fields
\begin{equation}
\label{eq:Willmore_eps}
\mathcal{W}_\eps(u) =
\begin{cases}
\displaystyle
\frac{1}{\eps} \int_\W \left( \eps \Lap u - \frac{W'(u)}{\eps}\right)^2 ~dx ~~\text{if } u \in W^{2,2}(\W) \\
+ \infty ~~\text{otherwise in } L^1(\W)
\end{cases}
\end{equation}
Note that the trace term $\Tra{\Meps} = -\eps \Lap u + \frac{W'(u)}{\eps}$ inside the square is the $L^2$ gradient of $\frac{\eps}{2} \ngd^2 + \frac{W(u)}{\eps}$ which appears in the Cahn-Hilliard phase-field \eqref{eq:CH_eps}, in the same way as the mean curvature vector is the $L^2$ gradient of the area functional.

The $\Gamma$-limsup was showed in \cite{bellettini_approssimazione_1993}, using the same recovery sequence as for the area functional. The $\Gamma$-liminf was studied under several conditions in \cite{bellettini_approximation_2005,moser_higher_2005} and completed in \cite{roger_modified_2006}. Together with the $\Gamma$-limsup, this resulted in the $\Gamma$-convergence on smooth points of the form $u = 2 \chi_E - 1$ where $E \subset \W$ and $\bord E \inter \W$ is $C^2$:
\begin{equation*}
\left[\Gamma(L^1(\W)) - \lim\limits_{\eps \to 0} \mathcal{W}_\eps\right]( 2\chi_E - 1) = \sigma ~\Will(\bord E \inter \W),
\end{equation*}
but with the additional assumption that the diffuse surface areas $A_\eps$ remain uniformly bounded.\newline

Finally, Bellettini and Mugnai \cite{bellettini_approximation_2010} extended the Willmore phase-field to approximate the complete Helfrich energy with
\begin{equation}
\label{eq:Helfrich_eps}
\mathcal{H}_\eps(u) =
\begin{cases}
\displaystyle
\int_\W \left[ \frac{\chi_b}{2 \eps} \left(\Tra{\Meps} - \eps \ngd H_0 \right)^2 \right. \\ 
\displaystyle ~~ +  \vphantom{\int_\W1}  \left.\frac{\chi_G}{2\eps} \left( (\Tra{\Meps})^2 - \Nor{\Meps}^2 \right) \right] ~dx ~~\text{if } u \in C^2(\W),\\
+ \infty ~~\text{otherwise in } L^1(\W).
\end{cases}
\end{equation}
Based on the previous results of R\"{o}ger and Sch\"{a}tzle \cite{roger_modified_2006}, and under the assumptions $H_0 = 0$ and $- \chi_b < \chi_G < 0$, they showed that the $\Gamma$-convergence holds on smooth points $u = 2 \chi_E - 1$ where $E \subset \W$ is open and $\bord E \inter \W$ is $C^2$:
\begin{equation}
\left[ \Gamma(L^1(\W)) - \lim\limits_{\eps \to 0^+} \mathcal{H}_\eps  \right](2 \chi_E - 1) = \sigma ~ \Hel(\bord E \inter \W),
\end{equation}
again using an additional uniform bound on the diffuse areas $A_\eps$.\newline

Our aim is precisely to generalize the Helfrich phase-field formula $\mathcal{H}_\eps$ further to approximate the curvature functional \eqref{eq:F}, and provide a computational framework to simulate shape textures. We are now ready to construct a new phase-field, for which we will study the $\Gamma$-limsup property.

\section{Construction of the phase-field}
\label{sec:model}

In this section, we generalize the Helfrich phase-field $\mathcal{H}_\eps$ in \eqref{eq:Helfrich_eps} to approximate the functional $\Fbf$ in \eqref{eq:F}, using the notations introduced in Section \ref{sec:classical_phase_fields}. We justify the construction with heuristic arguments, and show that the $\Gamma$-limsup is still satisfied, although we do not attempt to show the $\Gamma$-liminf.

\subsection{Diffuse curvatures and second fundamental form}
Let us notice that the diffuse expressions $\mathcal{W}_\eps$ and $\mathcal{H}_\eps$ in \eqref{eq:Willmore_eps} and \eqref{eq:Helfrich_eps} both rely on the trace and the norm of the matrix field $\Meps$ introduced in \eqref{eq:Meps}, which is related to the second fundamental forms and the curvatures of the level sets of $u$ as follows.
We define the \textit{diffuse second fundamental form} $\Beps$ as well as the \textit{diffuse mean and Gaussian curvatures} $\Heps$ and $\Keps$ using
\begin{align}
\Beps &= \frac{\Meps}{\eps \ngd} \label{eq:Beps} \\
\Heps &= \frac{\Tra{\Meps}}{\eps \ngd} \label{eq:Heps}\\
\Keps &= \frac{1}{2 \eps^2 \ngd^2} \left[ (\Tra{\Meps})^2 - \|\Meps\|^2 \right] \label{eq:Keps}
\end{align}
if $\grad u \neq 0$, and zero otherwise.
Informally, $\Beps \otimes \n_u$ at the point $x \in \W$ approximates the second fundamental form of the level surface $\{u = u(x)\}$ (well-defined if $\grad u \neq 0$ on this set). $\Heps \n_u$ approximates the mean curvature vector, with the convention that it points inwards for convex sets, and $\Keps$ approximates the Gaussian curvature.

Based on the relations \eqref{eq:HK_kakb} linking $(\ka,\kb)$ to $(H,K)$, we also introduce the \textit{diffuse principal curvatures}
\begin{align}
\kepsa &= \frac{\Heps + \sqrt{ \left((\Heps)^2 - 4 \Keps \right)^+}}{2} \label{eq:kepsa}\\
\kepsb &= \frac{\Heps - \sqrt{ \left((\Heps)^2 - 4 \Keps \right)^+}}{2} \label{eq:kepsb},
\end{align}
where we use the positive part $x^+ = \max(0,x)$.

It can be shown, using the implicit formulas summarized in \cite{goldman_curvature_2005}, that the expressions from \eqref{eq:Beps} to \eqref{eq:kepsb} coincide exactly with the second fundamental form and the respective curvatures of the level sets of $u$, in the special case where the function has a \textit{hyperbolic tangent profile}
\begin{equation} \label{eq:tanh_profile}
u = \tanh \left(\frac{\mathrm{dist}_{\bord E}}{\sqrt{2} \eps} \right),
\end{equation}
where $E \subset \W$ is an open set with smooth boundary $\bord E \inter \W \in \mathcal{C}^2$, and the \textit{signed distance} from $\bord E$, denoted by $\mathrm{dist}_{\bord E}$, is by convention positive on $E$ and negative on $\W \setminus \bar{E}$. In diffuse approximations, the $\tanh$ profile is optimal\footnote{It is the unique minimizer of the 1D version of the Cahn--Hilliard energy among increasing functions with limits $\pm 1$ at $\pm \infty$.}
and is generally used to construct the $\Gamma$-limsup recovery sequence.\newline

The presence of the positive part in \eqref{eq:kepsa} and \eqref{eq:kepsb} ensures that the square root term is still defined when
$(\Heps)^2 - 4 \Keps < 0$. This can happen, since for a general $u$, we have $(\Heps)^2 - 4 \Keps = \frac{2 \Nor{\Meps}^2 - (\Tra{\Meps})^2}{\eps^2 \ngd^2}$, and the numerator is $a^2 + b^2 + c^2 - 2 \, (ab + bc + ac)$ which possibly has negative values, where $a, b, c$ are the real eigenvalues of $\Meps$. However, if one of them is $0$, the numerator is a squared difference and the positive part is not useful. This is the case in particular for functions with $\tanh$ profile.

The expressions \eqref{eq:kepsa} and \eqref{eq:kepsb} can be reformulated with $\Meps$, by writing
\begin{align*}
\kepsa &= \frac{\Tra{\Meps} + \sqrt{ \left(2 \Nor{\Meps}^2 - (\Tra{\Meps})^2 \right)^+}}{2 \eps \ngd} \\
\kepsb &= \frac{\Tra{\Meps} - \sqrt{ \left(2 \Nor{\Meps}^2 - (\Tra{\Meps})^2 \right)^+}}{2 \eps \ngd}, \\
\end{align*}
if $\grad u \neq 0$, and zero otherwise. Note that $\Heps$ can be retrieved from their sum
\begin{equation*}
\kepsa + \kepsb = \Heps
\end{equation*}
contrarily to $\Keps$ since their product is
\begin{equation*}
\kepsa \kepsb = \frac{1}{4} ~ \left[ (\Heps)^2 - \left((\Heps)^2 - 4 \Keps \right)^+ \right] \leq \Keps.
\end{equation*}


\subsection{Phase-field construction}
\label{sec:Feps_expression}
Let $\abf = (a_{2,0},\, a_{1,1},\, a_{0,2},\, a_{1,0},\, a_{0,1},\, a_{0,0}) = (a_\alpha)_{|\alpha| \leq 2} \in \R^6$ be a vector of real coefficients.
The associated polynomial function is denoted by $p(x,y) = \sum_{|\alpha| \leq 2} a_\alpha (x, y)^\alpha$, so that $\Fbf(\S) = \int_\S p(\ka,\kb) ~dA $.
Consider the following expression,
\begin{equation*}
\label{eq:Eeps}
\Eeps(u) =  \int_\W p(\kepsa, \kepsb) \, \eps \ngd^2 ~dx.
\end{equation*}
The heuristic intuition behind is that, if $u$ has a tanh profile with transition parameter $\eps$ as in \eqref{eq:tanh_profile}, we can apply the co-area formula to obtain
\begin{align*}
\begin{split}
& \int_{-1}^{1} \left( \int_{ \{u = t\} } p(\kepsa, \kepsb) \, \eps \ngd ~d\hauss^2 \right) ~dt\\
&= \int_{-1}^{1} \sqrt{2 W(t)}  ~ \Fbf( \{u = t\} ) ~dt,
\end{split}
\end{align*}
where we use $\ngd = \frac{(1-u^2)}{\sqrt{2} \eps} \neq 0$ and $\frac{1-t^2}{\sqrt{2}} = \sqrt{2 W(t)}$.
This amounts to integrating the curvature functional $\Fbf$ over all the level surfaces $\S_t = \{u = t\}$ of the phase-field $u$, appropriately weighted so that the largest contributions are given by level sets close to $\{ u = 0\}$. As $\eps$ goes to zero, the level sets $\{u = t\}$ concentrate around $\bord E$. \newline

Still under the ansatz of $\tanh$ profile, $\Eeps$ can be developed in terms of the matrix field $\Meps$:
\begin{align*}\label{eq:tildeEeps}
\begin{split}
\widetilde{\Eeps}(u) &= \int_\W \left[\frac{a_{2,0} + a_{0,2} - a_{1,1}}{4 \eps}~ \left(2 \Nor{\Meps}^2 - (\Tra{\Meps})^2 \right)^+ \right. \\
&+ \frac{a_{2,0} + a_{0,2} + a_{1,1}}{4 \eps} ~(\Tra{\Meps})^2 \\
&+ \frac{a_{2,0} - a_{0,2}}{2 \eps} ~\Tra{\Meps} \sqrt{\left(2 \Nor{\Meps}^2 - (\Tra{\Meps})^2 \right)^+} \\
&+ \frac{a_{1,0} + a_{0,1} }{2} ~\ngd \Tra{\Meps} \\
&+ \frac{a_{1,0} - a_{0,1} }{2} ~\ngd \sqrt{(2 \Nor{\Meps}^2 - (\Tra{\Meps})^2 )^+} \\
&+ \left.  \vphantom{\int_\W1} a_{0,0} ~\eps \ngd^2 \right] ~ dx.
\end{split}
\end{align*}
For a general $u \in W^{2,2}(\W)$, the expressions of $\Eeps(u)$ and $\widetilde{\Eeps}(u)$ 
coincide if $u$ is such that
\begin{equation*}
\mathcal{L}^3 \left( \{\ngd = 0 \} \inter \{ \Meps \neq 0 \} \right)  = 0,
\end{equation*}
which is satisfied for functions with $\tanh$ profile.\newline

Finally, $\widetilde{\Eeps}(u)$ can be simplified further, by replacing the positive part in the first term outside the square root directly by $2 \Nor{\Meps}^2 - (\Tra{\Meps})^2$. As said earlier, this is true of the special $\tanh$ case, where $2 \Nor{\Meps}^2 - (\Tra{\Meps})^2 \geq 0$ is $(a - b)^2$, where $a$, $b$ and $0$ are the eigenvalues of $\Meps$. This leads to the final form, defined for any $u \in W^{2,2}(\W)$,
\definecolor{newblue}{rgb}{.1,0.15,.6}
{ \color{newblue}
\begin{align} \label{eq:Feps}
	\Feps(u)& = \int_\W \left[\frac{a_{2,0} + a_{0,2} - a_{1,1}}{2 \eps}~ \Nor{\Meps}^2 + \frac{a_{1,1}}{2 \eps} ~ (\Tra{\Meps})^2 \right. \nonumber \\
	&+ \frac{a_{2,0} - a_{0,2}}{2 \eps} ~\Tra{\Meps} \sqrt{(2 \Nor{\Meps}^2 - (\Tra{\Meps})^2 )^+}  \nonumber\\
	&+ \frac{a_{1,0} + a_{0,1} }{2} ~\ngd \Tra{\Meps} \\ 
	&+  \frac{a_{1,0} - a_{0,1} }{2} ~\ngd \sqrt{(2 \Nor{\Meps}^2 - (\Tra{\Meps})^2 )^+} \nonumber \\
	&+ \left. \vphantom{\int_\W1} a_{0,0} ~\eps \ngd^2 \right] ~ dx. \nonumber
\end{align}
}
This phase-field is devised to be a diffuse approximation of the sharp-interface functional $\Fbf$ \eqref{eq:F}, up to the multiplicative factor $\sigma$. 

\paragraph{Comparison with the Willmore and the Helfrich diffuse approximations.} 
It can be checked that the proposed formulation $\Feps$ is indeed a generalization of the previous formulas $\mathcal{W}_\eps$ and $\mathcal{H}_\eps$, by specifying the polynomial coefficients of the Willmore energy,
\[ p(\ka,\kb) = H^2  \quad \text{ with } \quad \abf = (1,2,1,0,0,0), \] and of the Helfrich energy,
\[ p(\ka,\kb) = \frac{\chi_b}{2} (H-H_0)^2 + \chi_G K \quad \text{ with } \] 
\[\abf = (\frac{\chi_b}{2}, \chi_b + \chi_G, \frac{\chi_b}{2}, - \chi_b H_0, - \chi_b H_0, \frac{\chi_b}{2} H_0^2).\]

\subsection{$\Gamma$-limsup property}
As explained in Section \ref{sec:gamma_background}, the approximation of the target functional $\Fbf$ by the sequence $(\Feps)_{\eps > 0}$ can be studied in the $\Gamma$-convergence framework. Here, we assert that the phase-field $\Feps$ satisfies the $\Gamma$-limsup property, thus extending the result of \cite{bellettini_approximation_2010}. 
\begin{theorem}[$\Gamma$-limsup inequality]
	\label{thm:gamma_limsup}
	Let $E \subset \W$ be a bounded open set, such that $\bord E \inter \W$ is of class $\mathcal{C}^2$. The functionals $\Feps$ and $\Fbf$ are defined as in \eqref{eq:Feps} and \eqref{eq:F}. Then there exists a sequence of functionals $(u_\eps)_{\eps > 0} \subset W^{2,2}(\W)$ such that
	\begin{align}
	&\lim\limits_{\eps \to 0^+} u_\eps = 2 \chi_E - 1 \text{ in } L^1(\W), \label{eq:limsupa} \\
	&\lim\limits_{\eps \to 0^+} \eps |\grad u_\eps|^2 \, \mathcal{L}^3_{| \, \W} = \sigma \, \hauss^2_{| \, \bord E} \text{ as Radon measures}, \label{eq:limsupb} \\
	&\lim\limits_{\eps \to 0^+} \Feps(u_\eps) = \sigma \, \Fbf(E) \label{eq:limsupc}
	\end{align}
\end{theorem}
The proof is given in the Appendix. It consists in showing that the recovery sequence constructed in \cite{bellettini_approximation_2010} still satisfies the theorem for our more general formulation $\Feps$. 
The first and third properties correspond to the existence of a recovery sequence. The second property loosely means that the measure whose density is $\eps |\grad u_\eps|^2$ in the 3D volumetric space concentrates into the measure induced by the area on the 2D surface. This intuition is in accordance with the way we constructed $\Feps$ (see beginning of Section \ref{sec:Feps_expression}).

However, whether or not the $\Gamma$-liminf holds still remains an open question, except in the special cases of the Willmore and the Helfrich energies with additional assumptions, as seen in Section \ref{sec:classical_phase_fields}. 
Yet, we believe that, even if the $\Gamma$-convergence could fail in general, this does not constitute a serious impediment to our phase-field expression being still of interest for generating shape textures.

\section{Simulations}
\label{sec:simulations}

In this section, we demonstrate that the phase-field $\Feps$ constructed in the previous section can generate a large range of shape textures. We describe \verb|curvatubes| in Algorithm \ref{algo:Hminus_flow}, and then show the results of four numerical experiments. The first one validates the approach by finding well-known Willmore minimizers. We then display a gallery of $10$ shape textures. The effect of smoothly varying the generation parameters is shown in the third experiment, with a bilinear interpolation between $4$ shape textures, layers, spheres, tubes, and sponges. Finally, $1000$ shape textures are generated with random parameters and visualized in an atlas with UMAP. The numerical codes are fully available at
\begin{center}
\verb|https://github.com/annasongmaths/curvatubes|~.
\end{center}
\subsection{Curvatubes}
\label{sec:algo_description}

Shape textures are generated by minimizing the phase-field energy $\Feps$ \eqref{eq:Feps} under a constraint of constant volume, with periodic boundary conditions. More exactly, given a random initialization of the phase-field $u$, we find a point of convergence with low energy of the so-called $H^{-1}$ flow\footnote{This is a gradient flow with respect to the $H^{-1}(\W)$ metric, where $H^{-1}(\W)$ is the dual of the space $H_0^1(\W)$, the closure of the set $C^\infty_c(\W)$ of smooth compactly-supported functions in $W^{1,2}(\W)$ \cite{evans_partial_2010,cowan_cahn-hilliard_2005}.}
\begin{equation*}
\dot{u} = \Lap \frac{\partial \Feps}{\partial u}.
\end{equation*}
The $H^{-1}$ flow is \textit{mass-preserving}, i.e., keeps constant the \textit{mass} of $u$, denoted by $\bar{u} := \frac{1}{\W} \int_\W u ~dx$. The preservation of mass approximately encodes a constraint of constant volume on the region enclosed by the surface $\{u = 0\}$, if the phase-field $u \simeq \pm 1$ is nearly constant inside and outside.\newline

The $H^{-1}$ flow can actually be expressed as a standard $L^2$ flow, by relying on the change of variable 
\begin{equation}
\label{eq:change_var}
u = \Div A + m_0,
\end{equation}
where $A : \W \to \R^3$ is a periodic vector field, and $m_0 \in \R$ is the desired value of the average $\bar{u}$. We then define an energy with respect to $A$,
\begin{equation*}
G_\eps(A) = \Feps(\Div A + m_0).
\end{equation*}
It can be checked that
\[ \frac{\partial G_\eps}{\partial A}(A) = - \grad \frac{\partial \Feps}{\partial u}(u), \]
in such a way that a $L^2$ flow on $A$ becomes a $H^{-1}$ flow on $u$:
\[ \dot{A} = - \frac{\partial G_\eps}{\partial A} \quad \Rightarrow \quad \dot{u} = \Lap \frac{\partial \Feps}{\partial u}\]
(provided that the derivatives in time and space of $A$ commute).
Therefore, the $H^{-1}$ flow on $u$ starting at $u_0 = \Div A_0 + m_0 $ can be solved as a usual $L^2$ flow on $A$. \newline

To generate shape textures, the variable $A$ is initialized as a random white noise vector field $A_0$ and we reach a point of convergence of the $L^2$ flow $\dot{A} = - \frac{\partial G_\eps}{\partial A}$ with Adam \cite{kingma_adam_2015,loshchilov_decoupled_2019}, a gradient-based stochastic optimization algorithm.
The change of variable \eqref{eq:change_var} allows us to benefit from the computation of the $L^2$ gradient $\frac{\partial G_\eps}{\partial A}$ by the automatic differentiation engine provided by PyTorch \cite{paszke_pytorch_2019}, combined with the efficiency of Adam. 

\begin{figure*}[h!]
	\centering
	\includegraphics[trim = .2cm 0 .2cm 0 , clip, width=\linewidth]{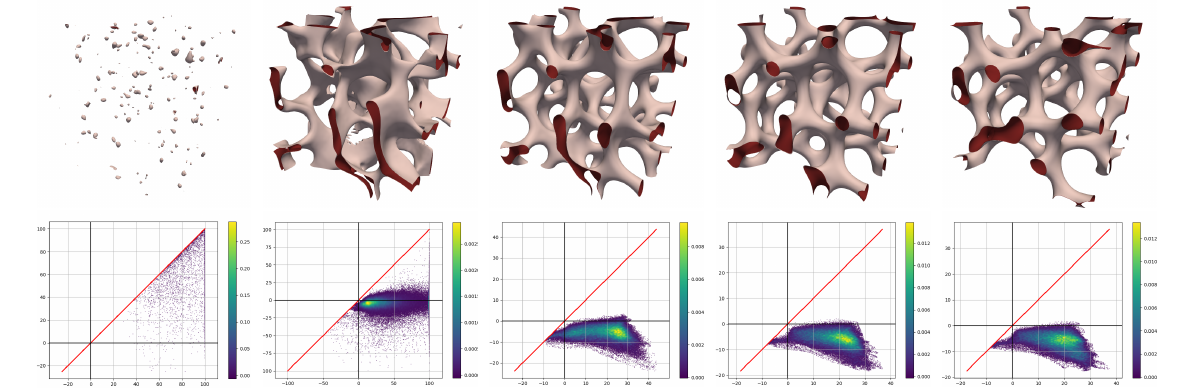} 
	\caption{\small\textbf{Evolution of the zero level set of $u$ during the $H^{-1}$ flow}, pictured at iterations $10, 300, 1000, 4000, 8000$, with the generation parameters $\abf = (1, 2.4, 9, 30, 170, -195)$ and $m_0 = -0.66$. Top row: the shape textures. Bottom row: their corresponding curvature diagrams. In the early iterations, the zero level set is not smoothly defined and only few values of $u$ are above zero, but their average remains $m_0$.}
	\label{fig:cvg_1}
\end{figure*}

\begin{figure}[h!]
	\centering
	\includegraphics[trim = 0 0 0 0 , clip, width=\linewidth]{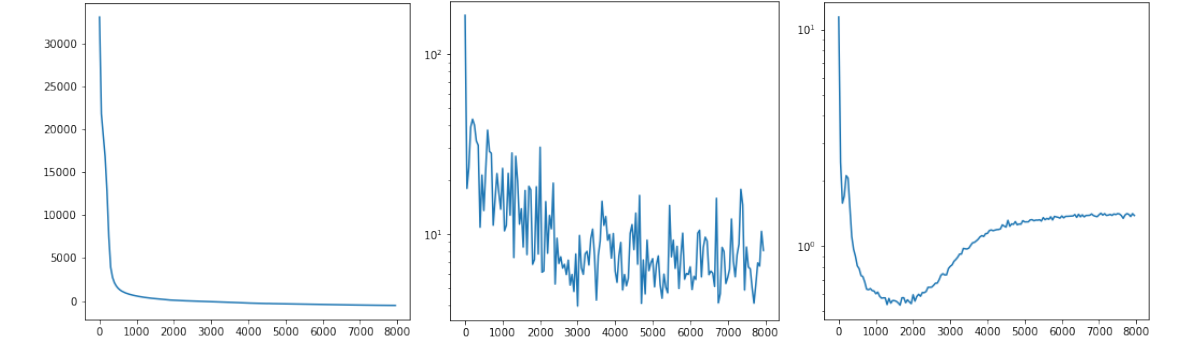} 
	\caption{\small\textbf{Corresponding loss curves (see Figure \ref{fig:cvg_1}).} From left to right: the loss $\Feps(u)$, the maximal value of  $|\frac{\partial G_\eps}{\partial A}|$ and the average $\frac{1}{|\W|} \int_\W |\frac{\partial G_\eps}{\partial A}|$, as $u$ evolves along the iterations.}
	\label{fig:cvg_1_curves}
\end{figure}

The generation model is summarized in Algorithm \ref{algo:Hminus_flow}.
It takes as inputs the initialization $A_0$, the coefficients $\abf$ and the mass $m_0$. After convergence, the output shape texture is defined as the level surface $\{u = 0\}$ of the final phase-field $u = \Div A + m_0$. We color it in beige, and show the level set $\{ u = 0.05 \}$ in dark red to enhance the visualization.
An example of flow is given in Figure \ref{fig:cvg_1}, with the corresponding loss curves in Figure \ref{fig:cvg_1_curves}.

\begin{algorithm}
	\small
	\caption{\small\textsc{Curvatubes}: generate shape textures of optimal curvature energy $\Feps$, using a mass-preserving $H^{-1}$ flow on the phase-field $u$ and periodic boundary conditions}
	\label{algo:Hminus_flow}
	\begin{algorithmic}[1]
		\Procedure{Curvatubes}{$A_0; \abf, m_0$} \newline
		\textbf{Initialization:} random vector field $A_0$ \newline
		\textbf{Generation parameters:} coefficients $\abf = (a_{2,0},\, a_{1,1},\, a_{0,2},  \, a_{1,0},\, a_{0,1},\, a_{0,0})$, mass $m_0 \in (-1,1)$ \newline
		\textbf{Energy:} phase-field energy $\Feps$ (see \eqref{eq:Feps}) to approximate $\int_\S ( \sum_{|\alpha| \leq 2} a_\alpha (\ka,\kb)^\alpha) ~dA$ \newline
		\textbf{Other parameters:} phase transition parameter $\eps > 0$, internal parameters for Adam (learning rate, betas, weight decay), number of iterations $T$, Gaussian kernel of size $\sigma_k$ \newline
		\textbf{Outputs:} phase-field $u$ and surface $\S = \{ u = 0 \}$ \newline

		\State{$A \leftarrow A_0$}
		\Comment{initialization of the vector field}	
		
		\For{$t = 1,...,T$}			 
		\State{$u \leftarrow \Div A + m_0$} \Comment{change of variable \eqref{eq:change_var}}
		\State{$u \leftarrow k \ast u$} \Comment{small blur to avoid artifacts}
		\State{Set $G_\eps(A) := \Feps(u)$}
		\State{Compute $\frac{\partial G_\eps}{\partial A}$} \Comment{PyTorch autograd}
		\State{Update moments of the gradients} \Comment{Adam} 
		\State{Update $A$ with one step of $\dot{A} = - \frac{\partial G_\eps}{\partial A} $}  \Comment{Adam}
		\EndFor	
		
		\State{$\S \leftarrow \{ u = 0\}$}
		\Comment{visualize the shape texture}
		\EndProcedure
		
	\end{algorithmic}
\end{algorithm}

\paragraph{Implementation details.}

The domain $\W$ is assimilated to a grid of size $100 \times 100 \times 100$ pixels with a fixed sampling step $\Delta x = 0.01$. We take $\eps = 0.02$, unless specified otherwise.
The phase-field $u$ and the vector field $A$ are encoded as matrices whose coefficients specify the sampled values. 

The discrete energies $\Feps$ (resp. $G_\eps$) are symbolically defined by a succession of elementary operations on $u$ (resp. $A$), before being differentiated automatically by PyTorch. 
In particular, the integral is encoded as a finite sum, while the differential operations $\grad u$, $\mathrm{Hess}~ u$, and $\Div A$, are computed as classical finite differences that take into account the periodicity of the problem. To prevent the formation of artifacts, we apply a Gaussian blur $k$ with a small deviation (typically $\sigma_k = 2$ pixels) to the phase-field $u$, before computing the finite differences. 
The norm of the gradient $\ngd$ is modified by a small offset $\xi = 10^{-6}$, as in $\sqrt{\ngd^2 + \xi^2}$ or $1/\sqrt{\ngd^2 + \xi^2}$, to prevent non-differentiability at zero and division by zero. The positive part function $x^+$ appearing in $\Feps$ is approximated by a smooth function $x^+ \simeq \xi \log(1 + \mathrm{e}^{x / \xi})$.

With Adam, the step size and direction at each point are computed in an adaptive way, by taking into account the past history of the gradients to estimate their first and second moments. In the simulations presented thereafter, Adam was run with a learning rate $lr = 0.001$, $betas = (0.9, 0.999)$, and no weight decay. We stopped the algorithm typically after $8000$ iterations, as the convergence was estimated to be reached, which induced up to $150$ seconds of computation time per shape with a simulation domain of size $100 \times 100 \times 100$ pixels. 

\paragraph{Generation parameters, shape textures, curvature diagrams.}

The generation of shape textures relies on the principle that a generation parameter vector $(\abf, m_0)$ should consistently correspond to a single shape texture, across different white noise initializations of $A_0$. We ``measure" the texture of a shape defined by a surface $\S$ through its \textit{curvature diagram}, which represents the distribution of the curvatures $(\ka,\kb)$ on $\S$ (see bottom row of Figure \ref{fig:cvg_1} for instance). More precisely, we are interested in the law of the random variable $(\ka,\kb)$ defined by
\begin{align*}
\mathbb{P}\left[ (\ka,\kb) \in \mathcal{B} \right] &= \frac{1}{\text{Area}(\S)} \int_\S \chi_{(\ka,\kb) \in \mathcal{B}} ~d\hauss^2  \\
&= \frac{\text{Area}( \{ x \in \S ~|~ (\ka,\kb)(x) \in \mathcal{B} \} )}{\text{Area}(\S)}, 
\end{align*}
for a Borel set $\mathcal{B}$ of $\R^2$. 

The curvature diagram is an indicator of the local behavior of the surface. For a perfect sphere of radius $R$, it should be a unit Dirac mass $\delta_{(\frac{1}{R},\frac{1}{R})}$ sitting on the $\{ y = x, \,x > 0 \}$ half-diagonal, since $\ka = \kb = \frac{1}{R}$ everywhere on a sphere. Likewise, a cylinder of base radius $r$ should have its diagram reduced to $\delta_{(\frac{1}{r}, 0)}$ on the horizontal half-line $\{ y = 0, \,x > 0 \}$; a plane would be represented as $\delta_{(0,0)}$; finally, a Dirac mass such as $\delta_{(1,-1)}$ should correspond to a sponge-like shape. Note that, per definition of the curvatures $\ka \geq \kb$, the distribution is contained in the lower mid-plane $\{ y \leq x \}$.

To obtain a curvature diagram, we first extract the 2D mesh of the surface $\S = \{ u = 0 \}$ from the 3D volume $u$, by using the marching cubes algorithm \cite{lewiner_efficient_2003}. Then the diffuse curvatures $(\kepsa, \kepsb)$ (see \eqref{eq:kepsa} and \eqref{eq:kepsb}) are interpolated at the barycenter of each cell of the mesh. Their values are of importance proportional to the area of the cell, resulting in a weighted point cloud 
\begin{equation} \label{eq:curva_diagram}
\sum_{\text{cells } c} \text{Area}\,[c] ~\delta_{(\kepsa, \kepsb)\,[c]}
\end{equation}
which we plot in the curvature diagram. In our simulations, the shapes are rarely perfectly spherical, cylindrical, or flat, so that the distribution is dispersed rather than concentrated into a single Dirac mass. To ease the visualization of the curvature diagrams, the identity diagonal $\{y = x\}$ is enhanced as a red line and the values truncated between $-100$ and $100$.

We compare curvature diagrams with each other using the Wasserstein (or Earth Mover's) distance. We approximate this quantity with the regularized Sinkhorn algorithm of the \verb|geomloss| module \cite{feydy_interpolating_2019}, with the parameters $p = 2$, $blur = 1$ and $reach = 20$.
As detailed in \cite{feydy_geometric_2020}, these correspond to the resolution of an unbalanced transport problem \cite{sejourne_sinkhorn_2019} using a ground cost function of 
$\text{C}(x,y)=\tfrac{1}{2}\|x-y\|^2$, with a transport plan that is blurred at resolution of 1 and with a maximum transport distance of the order of 20.

\begin{figure*}[h!]
	\centering
	\includegraphics[trim = 0 0 0 0 , clip, width=\linewidth]{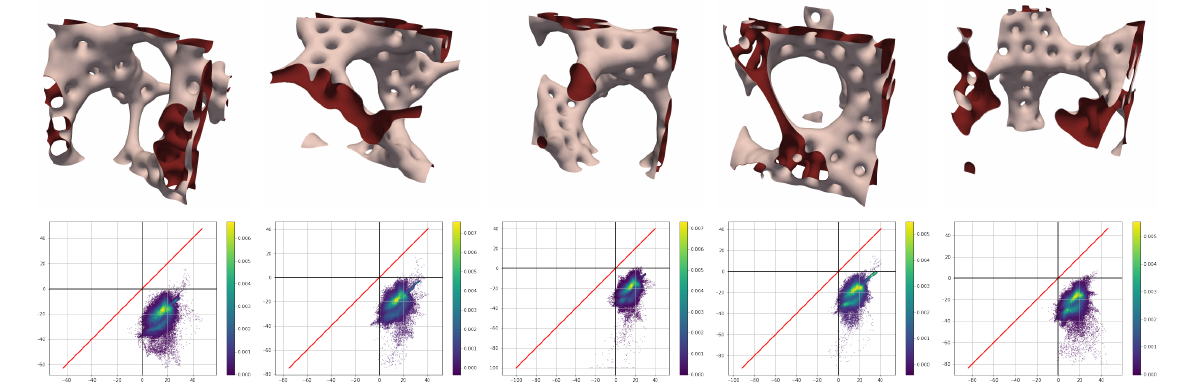} 
	\caption{\small \textbf{Same generation parameters, different initializations.} Using the same generation parameters $\abf = (1, -0.35, 1.02, -40, 100, 1600) $ and $m_0 = -0.69$, we start from five random initializations for $A_0$. Top row: the shape textures. Bottom row: their corresponding curvature diagrams. The five shapes are visually similar at a mesoscopic scale, but dissimilar at the macroscopic scale. The diagrams and their pairwise Wasserstein distances show that the curvature statistics are not changing much when feeding the same generation parameters to the algorithm. Texture in shapes hence seems to be well captured by curvature diagrams, and to be consistent with the generation parameters at a mesoscopic scale.}
	\label{fig:var_1}
\end{figure*}

Figure \ref{fig:cvg_1} shows that the curvature diagrams of the evolving level surface also converge towards a final diagram. In Figure \ref{fig:var_1}, we check that for a single generation parameter value $(\abf,m_0)$, five different initializations still give similar curvature diagrams. We found that the mean pairwise Wasserstein distance between them was only $10.7 \%$ of the mean pairwise distance measured between $1000$ random shapes (see Experiment 4). Therefore, curvature diagrams and generation parameters seem to capture well the notion of shape texture.

\subsection{Experiment 1: validation with the Willmore flow and known minimizers of genus 0, 1, and 2}

The algorithmic framework, which combines automatic differentiation and control of gradient flows by external optimizers, is validated in the fundamental special case of the $L^2$ Willmore flow (see Algorithm \ref{algo:L2_flow}, with the parameters $\abf = (1,2,1,0,0,0)$, and with replicate boundary conditions\footnote{The replicate padding on $u$ corresponds to the assumption that its gradient is orthogonal to the domain boundary, i.e., $\grad u \cdot \n_{\bord \W} = 0$.} ). Let us recall from Section \ref{sec:classical_phase_fields} that the Willmore phase-field energy writes $\Feps = \frac{1}{\eps} \int_\W ( \eps \Lap u - \frac{W'(u)}{\eps})^2 dx$ and approximates $\sigma \int_\S H^2 \, dA$. We numerically check that the simulated Willmore flow converges towards known global minimizers of fixed genus $0$, $1$, and $2$, by initializing the flow near them. The gradient descent is controlled by the L-BFGS optimizer, as it was empirically found to converge faster than Adam. L-BFGS approximates the BFGS algorithm, a quasi-Newton method that combines a line search to an estimation of the Hessian of the loss \cite{fletcher_practical_1987,kelley_iterative_1999}.

The Willmore minimizers of genus $0$ are spheres of any radius \cite{willmore_riemannian_1996}, which achieve the minimal value\footnote{We have to multiply the conventional values by $4$, as $H$ is defined as the (real) mean of the curvatures $\frac{\ka + \kb}{2}$ in other works.}
\[ \int_{\S} H^2 ~dA = \int_\S \left(\ka + \kb \right)^2 ~dA = 4 \times 4 \pi.\]
For surfaces of genus $1$, \cite{marques_min-max_2014} proved that the minimal value $4 \times 2 \pi^2$ is achieved by the Clifford torus (up to conformal transformations), defined by a special ratio $\frac{1}{\sqrt{2}}$ between the radius of the generating circle and the distance to the axis of revolution.
However, the proof for genus $\geq 2$ has still not been completed, although several conjectures have been proposed. It has been shown that the minimum Willmore energy among all (orientable closed) surfaces of genus $g$ is less than $4 \times 8 \pi$, and converges to this value as the genus $g \to \infty$ \cite{kuwert_large_2010}. The Lawson surfaces have also been conjectured to be the minimizers for a given genus (up to conformal transformations) \cite{kusner_comparison_1989,hsu_minimizing_1992}. \newline

\begin{algorithm}
	\small
	\caption{\small\textsc{$L^2$ flow}: follow the gradient flow of the phase-field energy $\Feps$ until convergence, with given initialization and replicate boundary conditions}
	\label{algo:L2_flow}
	\begin{algorithmic}[1]
		\Procedure{$L^2$ flow}{$u_0; \abf$} \newline
		\textbf{Initialization:} a phase-field $u_0$ \newline
		\textbf{Generation parameters:} coefficients $\abf = (a_{2,0},\, a_{1,1},\, a_{0,2},\, a_{1,0},\, a_{0,1},\, a_{0,0})$ \newline
		\textbf{Energy:} phase-field energy $\Feps$ (see \eqref{eq:Feps}) to approximate $\int_\S ( \sum_{|\alpha| \leq 2} a_\alpha (\ka,\kb)^\alpha) ~dA$ \newline
		\textbf{Other parameters:} phase transition parameter $\eps > 0$, internal parameters for L-BFGS (learning rate, history size, line search function, maximal number of iterations in a line search), number of iterations $T$, Gaussian kernel of size $\sigma_k$ \newline
		\textbf{Outputs:} phase-field $u$ and surface $\S = \{ u = 0 \}$ \newline
		
		\State{$u \leftarrow u_0$}
		\Comment{initialization of the phase-field}	
		
		\For{$t = 1,...,T$}			 
		\State{$u \leftarrow k \ast u$} \Comment{small blur to avoid artifacts}
		\State{Compute $\frac{\partial \Feps}{\partial u}$} \Comment{PyTorch autograd}
		\State{Update $u$ with one step of $\dot{u} = - \frac{\partial \Feps}{\partial u} $}  \Comment{L-BFGS}
		\EndFor	
		
		\State{$\S \leftarrow \{ u = 0\}$}
		\Comment{Visualize the shape}
		\EndProcedure
		
	\end{algorithmic}
\end{algorithm}

\begin{figure}[h!]
	\centering
	\includegraphics[trim = 18cm 5cm 18cm 7cm, clip, width=.2\linewidth]{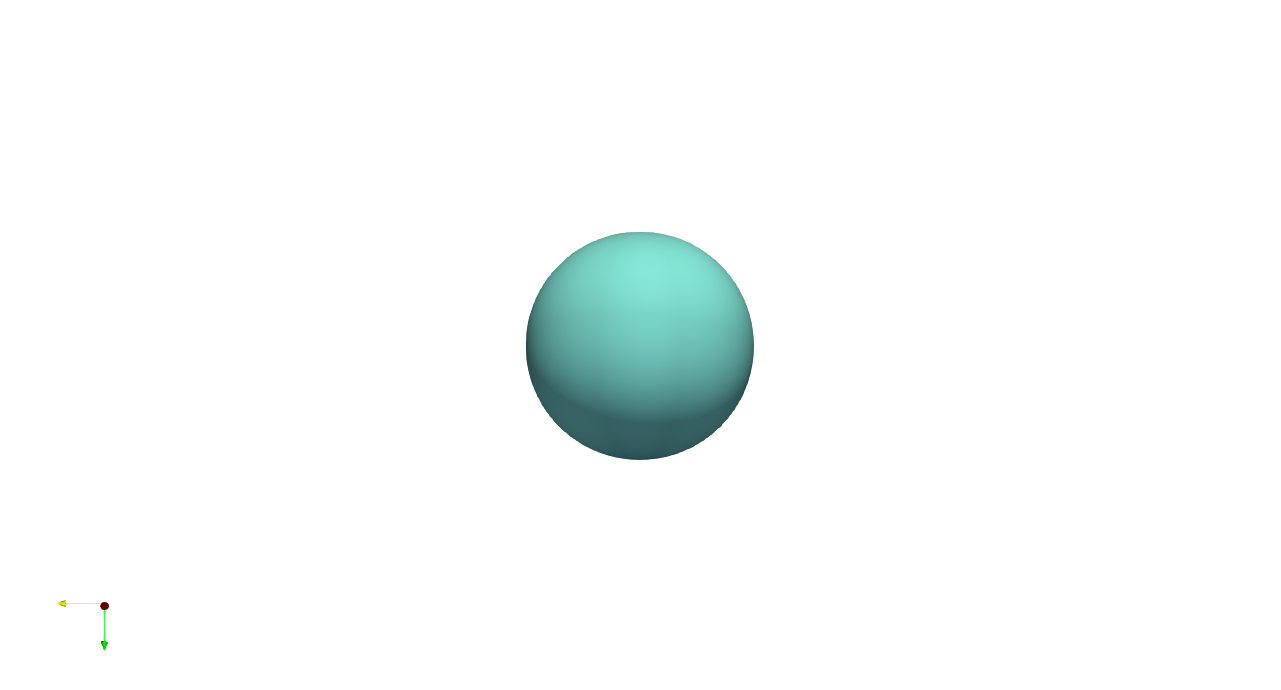} 
	\caption{\small\textbf{Willmore minimizer of genus $0$.} The flow is stationary on the sphere. The diffuse Willmore energy and the diffuse area are close to their mathematical values.}
	\label{fig:Willmore_genus0}
\end{figure}

\begin{figure}[h!]
	\centering
	\includegraphics[trim = 0 0 0 0, clip, width=.8\linewidth]{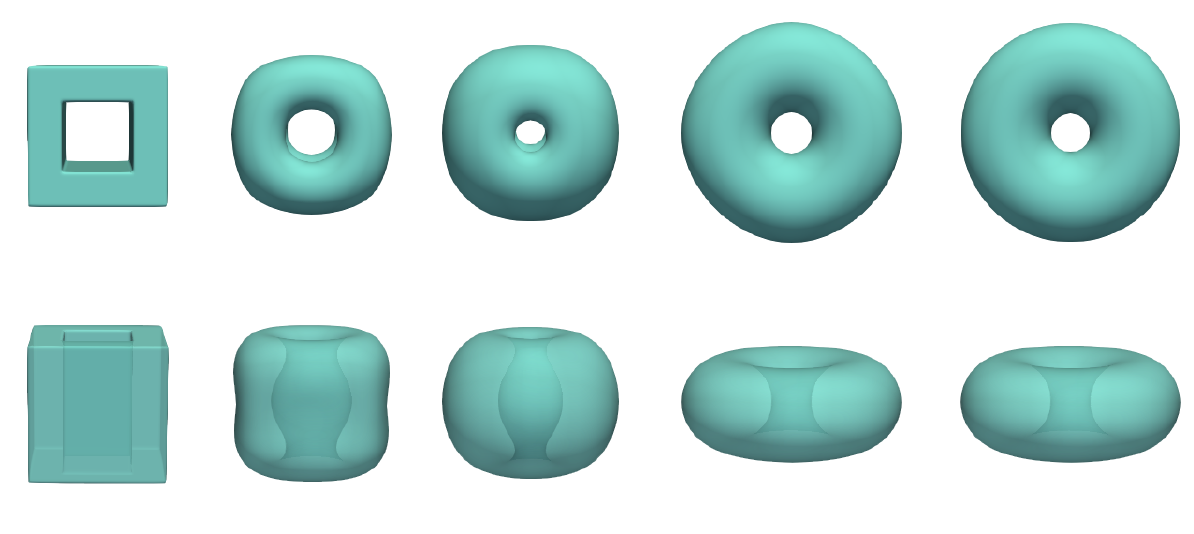} 
	\caption{\small\textbf{Willmore flow towards minimizer of genus $1$.} The flow converges to a Clifford torus. The ratio of the torus and the diffuse Willmore energy are close to their mathematical values.}
	\label{fig:Willmore_genus1}
\end{figure}

\begin{figure}[h!]
	\centering
	\includegraphics[trim = 0 0 0 0, clip, width=\linewidth]{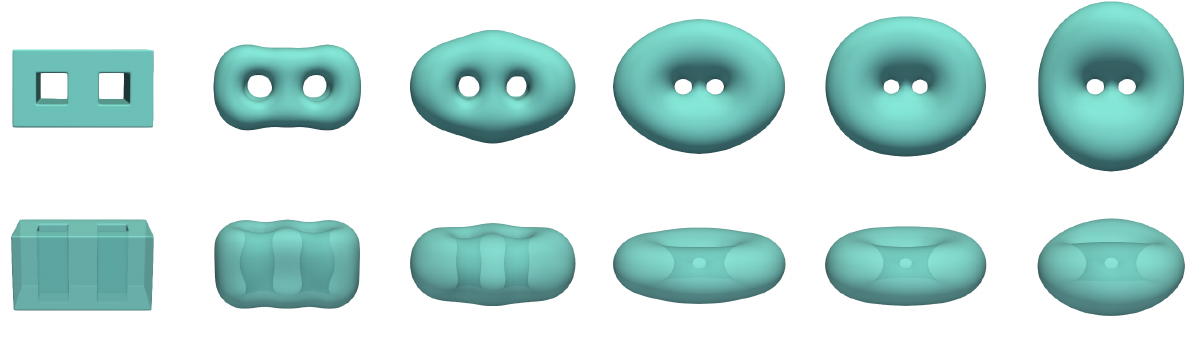} 
	\caption{\small\textbf{Willmore flow towards minimizer of genus $2$.} The flow converges to a Lawson-like surface, as expected from conjectures.}
	\label{fig:Willmore_genus2}
\end{figure}

We compare the final value $\frac{\Feps}{\sigma}$ to the minimal values mentioned above, where $\sigma = \frac{4}{3 \sqrt{2}}$ is the constant introduced in Section \ref{sec:background}.
We correctly find that the flow is stationary on spheres (see Figure \ref{fig:Willmore_genus0}).
The energy $\frac{\Feps}{\sigma}$ deviates from $4 \times 4 \pi$ with only $0.3 \%$ of relative error. We also check that the value of the Cahn-Hilliard energy $\int_\W (\frac{\eps}{2} \ngd^2 + \frac{W(u)}{\eps}) ~dA$, divided by $\sigma$, is close to the area of the sphere, with $0.5 \%$ of relative error.
When departing from a holed cube, the flow converges to a Clifford torus with a characterizing ratio close to $\frac{1}{\sqrt{2}}$ up to a relative error of $3\%$. We also find $\frac{\Feps}{\sigma} \simeq 4 \times 2 \pi^2$ up to a relative error of $0.4 \%$.
Finally, starting from two holed cubes glued together, the flow converges to a surface resembling a Lawson surface of genus $2$, with minimal value $\frac{\Feps}{4 \sigma} \simeq 22.30$, which has a relative difference of $1.8 \%$ compared to the value found in \cite{hsu_minimizing_1992}.
In Figures \ref{fig:Willmore_genus0}, \ref{fig:Willmore_genus1}, and \ref{fig:Willmore_genus2}, the parameters are $\eps = 0.04$, $\abf = (1,2,1,0,0,0)$ and $\sigma_k = 1$ pixel. The grid size is $200 \times 200 \times 200$ in the first two simulations and $200 \times 300 \times 400$ in the third one. L-BFGS was run with a learning rate $lr = 1$, a history size $hs = 10$, and maximum $20$ iterations in a line search.

\subsection{Experiment 2: a gallery of ten shape textures}

\begin{landscape}
	\begin{figure}[t]
		\centering
		\includegraphics[trim = 0 2cm 0 0, clip, page = 1, width=\linewidth]{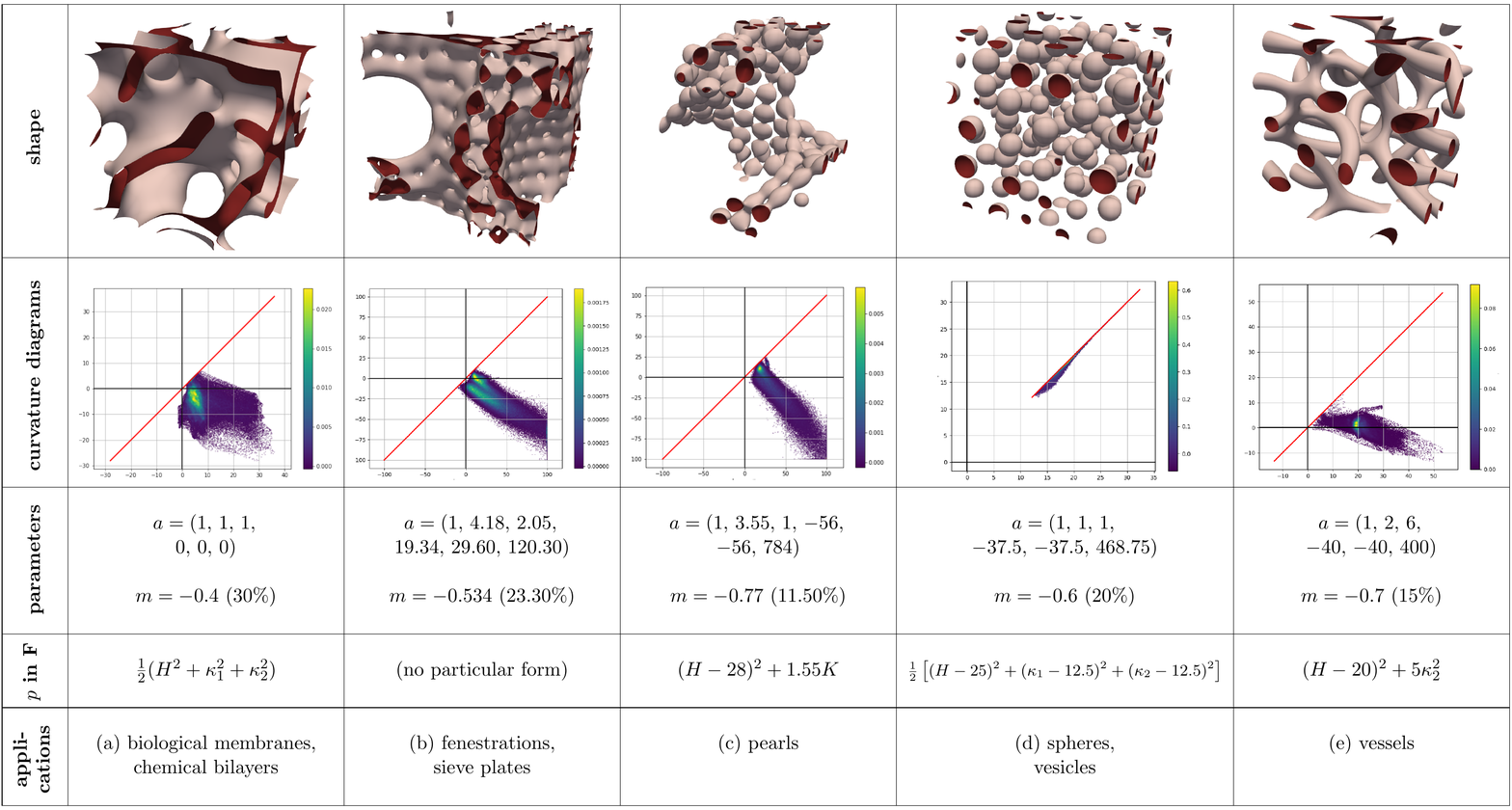}
		\caption{\small\textbf{Five shape textures (a, b, c, d, e).} We indicate the cases where the polynomial in the curvature energy has the form \eqref{eq:redundant}. }
		\label{fig:textures_a}
	\end{figure}
\end{landscape}

\begin{landscape}
	\begin{figure}[t]
		\centering
		\includegraphics[trim = 0 2cm 0 0, clip, page = 2, width=\linewidth]{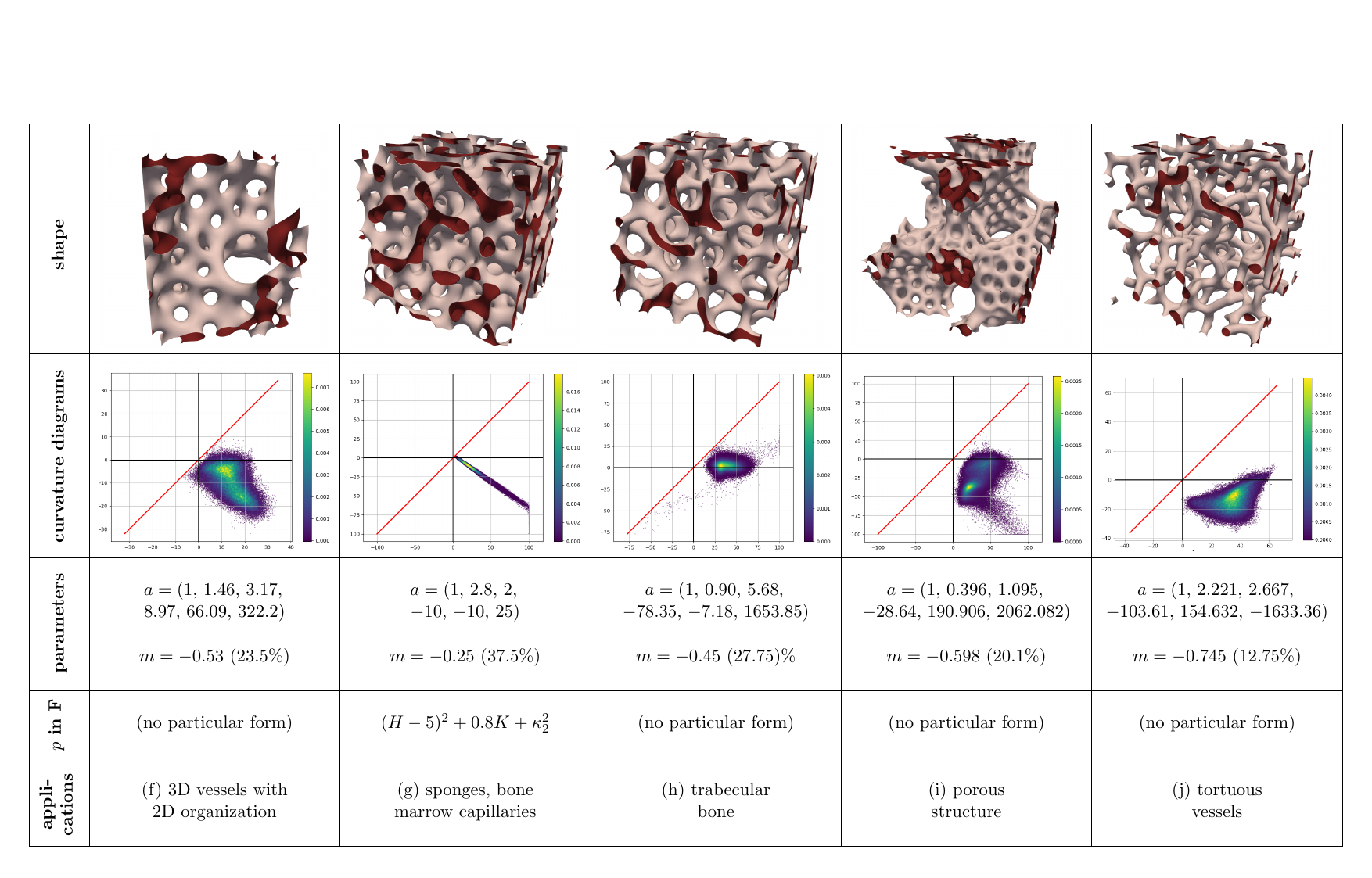}
		\caption{\small\textbf{Five shape textures (f, g, h, i, j).} We indicate the cases where the polynomial in the curvature energy has the form \eqref{eq:redundant}. }
		\label{fig:textures_b}
	\end{figure}
\end{landscape}

In Figures \ref{fig:textures_a} and \ref{fig:textures_b}, we show ten shape textures generated with Algorithm \ref{algo:Hminus_flow}, and visualize the surfaces together with their curvature diagrams. The coefficients $\abf$ and the mass $m_0 = \frac{1}{|\W|} \int_\W u$ used for the simulations are specified in the tables. The mass is also expressed in percentage of volume enclosed by the surface compared to the total volume of the domain, approximated by $\frac{m_0 + 1}{2}$ since $u \simeq \pm 1$.\newline

In practice, we find that the non-reduced polynomial expression
\begin{equation} \label{eq:redundant}
h_2 (H-H_0)^2 + k_1 K + \alpha (\ka - \ka^{0})^2 + \beta (\kb - \kb^0)^2
\end{equation} 
induces a choice of parameters $\bbf = (h_2, H_0, k_1, \alpha, \ka^0, \beta, \kb^0)$ that are more interpretable than the coefficients $\abf$ in the reduced form $\sum_{|\alpha| \leq 2} a_\alpha (\ka,\kb)^\alpha$. Using this formulation \eqref{eq:redundant}, it is easier to find reasonable values for which the energy leads to various tubular textures, without resulting in badly-converged phase-fields with no zero level set, or shapes broken into small fragments. The shapes were hence generated either by choosing $\bbf$ manually, or by selecting $\abf$ among random values from Experiment 4 that resulted in interesting shapes. \newline

We found that \verb|curvatubes| is able to generate very different shape textures (see Figures \ref{fig:textures_a} and \ref{fig:textures_b}, in which they are indexed by letters). Some shapes, such as (d), (e), (g), and (h) are smooth and spatially homogeneous in terms of visual aspect. Other shapes, such as (b), (c), (f), and especially (i), seem to possess a multi-scale texture or ``meta-texture" that makes them appear as spatially heterogeneous and anisotropic. The surface can be piecewise-smooth only, as in (c). Tubules are not necessarily smoothly turning cylinders, but can have some tortuosity as in (j). Finally, note that (h) combines flat regions and tubules, in a similar way to trabecular bone.




\subsection{Experiment 3: bilinear interpolation between four shape textures}

\begin{figure}[t]
	\centering
	\includegraphics[trim = 0 0 0 0, clip, width=\linewidth]{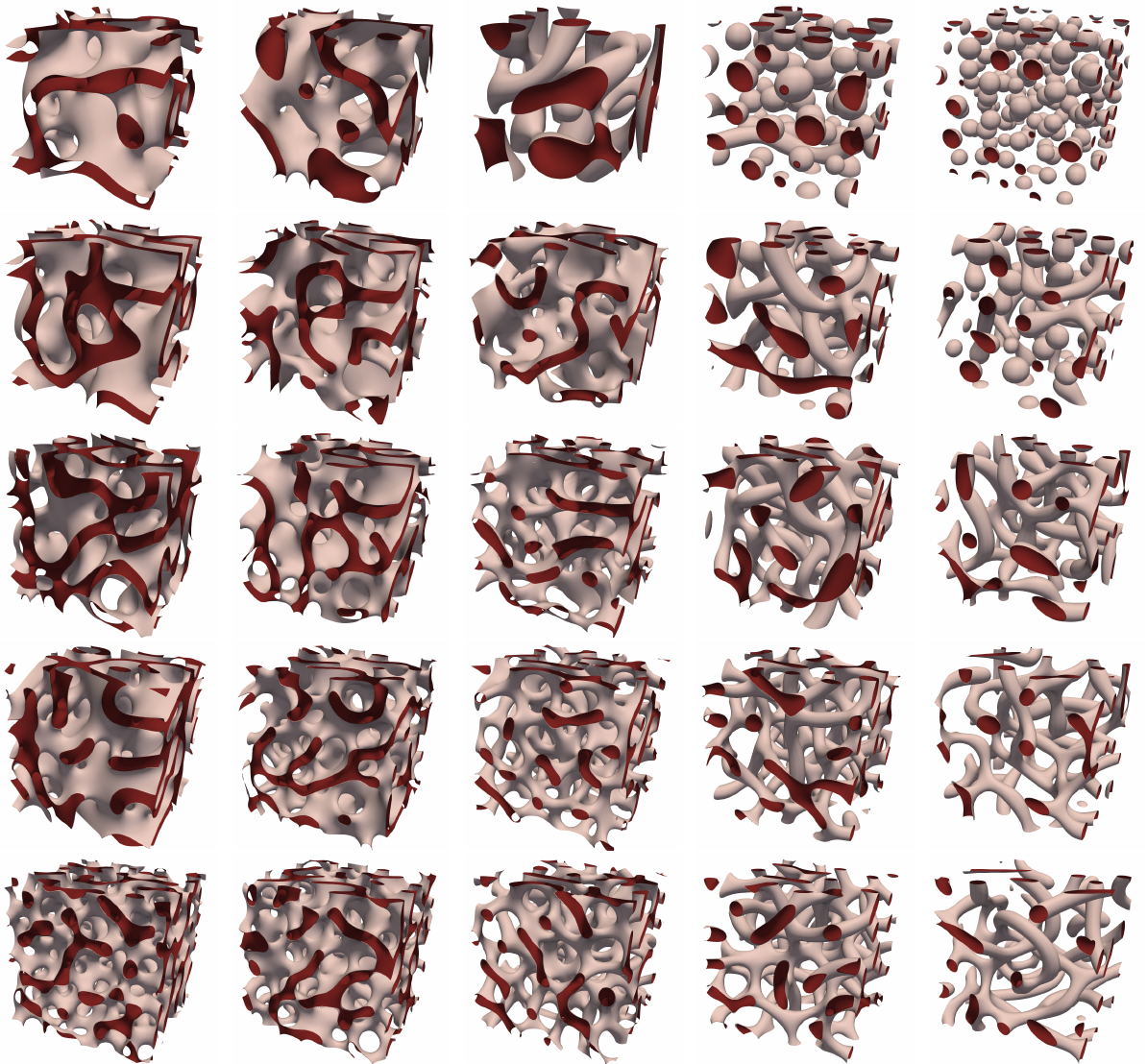} 
	\caption{\small \textbf{Bilinear interpolation between layers, spheres, tubes, and sponges.} Using the same initialization $A_0$ and interpolating between four reference generation parameters gives rise to a continuum of shape textures.}
	\label{fig:bili_shapes}
\end{figure}

In Figure \ref{fig:bili_shapes}, we illustrate how continuously varying the generation parameters $(\abf, m_0)$ impacts on the morphologies, by interpolating the parameters of four shape textures: layers, spheres, tubes, and sponges. The respective values can be found in Table \ref{tab:four_shapes}. All simulations are run by starting from the same initialization $A_0$.

\begin{table*}[t]
	\footnotesize
	\centering 
	\makebox[\textwidth][c]{
		\begin{tabular}{|c|c|c|c|c|}
			\hline
			shape texture & $\abf$ & $\bbf$ & $p$ in energy $\Fbf$ & $m_0$ \\[.1em]
			\hline
			layers & $(2,2,2,0,0,0)$ & $(1,0,0,1,0,1,0)$ & $H^2 + \ka^2 + \kb^2 $ & $-0.3 ~(35\%)$ \\
			spheres & $(2, 2, 2, -75, -75, 937.5)$ & $(1,2/R,0,1,1/R,1,1/R)$, $R = 0.08$ & 
			$(H-2/R)^2 + (\ka - 1/R)^2 + (\kb - 1/R)^2$ & $-0.6 ~(20\%)$ \\
			tubes & $(2, 2, 11, -100, -50, 1250)$ & $(1,1/r,0,1,1/r,10,0)$, $r = 0.04$ & $(H-1/r)^2 + (\ka-1/r)^2 + 10 \kb^2 $& $-0.7 ~(15\%)$ \\
			sponges & $(1, 2.8, 2, -10, -10, 25)$ & $(1,5,0.8,0,0,1,0)$ & $(H-5)^2 + 0.8 K + \kb^2 $ & $-0.25 ~(37.5\%)$ \\
			\hline
		\end{tabular}
	}
	\caption{\small\textbf{Generation parameters of four reference shape textures in Experiment 3} (see Figures \ref{fig:bili_shapes} and \ref{fig:bili_curvas}). The vector $\bbf$ parameterizes the polyomial expression \eqref{eq:redundant}. }
	\label{tab:four_shapes}
\end{table*}

\begin{figure}[t]
	\centering
	\includegraphics[trim = 0 0 0 0, clip, width=\linewidth]{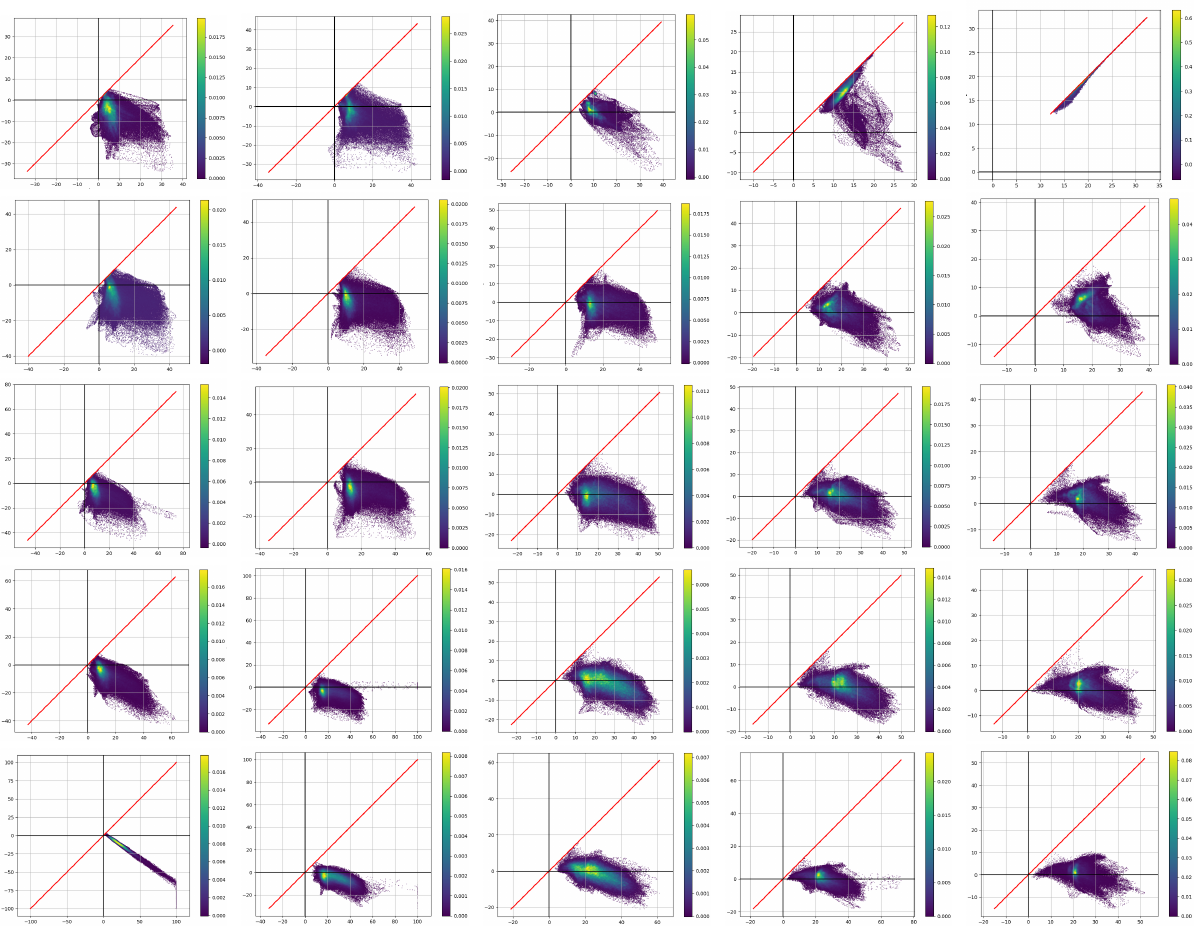} 
	\caption{\small \textbf{Corresponding curvature diagrams (see Figure \ref{fig:bili_shapes}).} The bilinear interpolation of generation parameters is also reflected into the curvature diagrams as a continuous evolution. Please note that the curvature distributions of the four reference shapes concentrate around $(0,0)$ for layers, the identity half-line $\{ y = x, x > 0 \}$ for spheres, the horizontal half-axis $\{ y = 0, x > 0 \}$ for tubes, and a half-line $\{ y = -a \,x, x > 0\}$ with $a > 0$ for sponges. As explained in Section \ref{sec:algo_description}, this is expected from curvature diagrams.}
	\label{fig:bili_curvas}
\end{figure}

We can see that the morphology is smoothly changing throughout the figure: for instance, the transition between spheres and tubes is characterized by tubes terminated on one side by end-caps, while layers increase in proportion compared to tubules when approaching the top left corner. The curvature diagrams are displayed in Figure \ref{fig:bili_curvas}, and they are quite continuously evolving as we change the generation parameters. The four diagrams at the corners reflect well the typical curvature distribution of layers, spheres, tubes, and sponges, as expected (see Section \ref{sec:algo_description}).


\subsection{Experiment 4: generation of 1000 shapes viewed in UMAP}

\begin{figure*}[t]
	\centering
	\includegraphics[trim = .1cm .1cm .1cm .1cm, clip, width=1\linewidth]{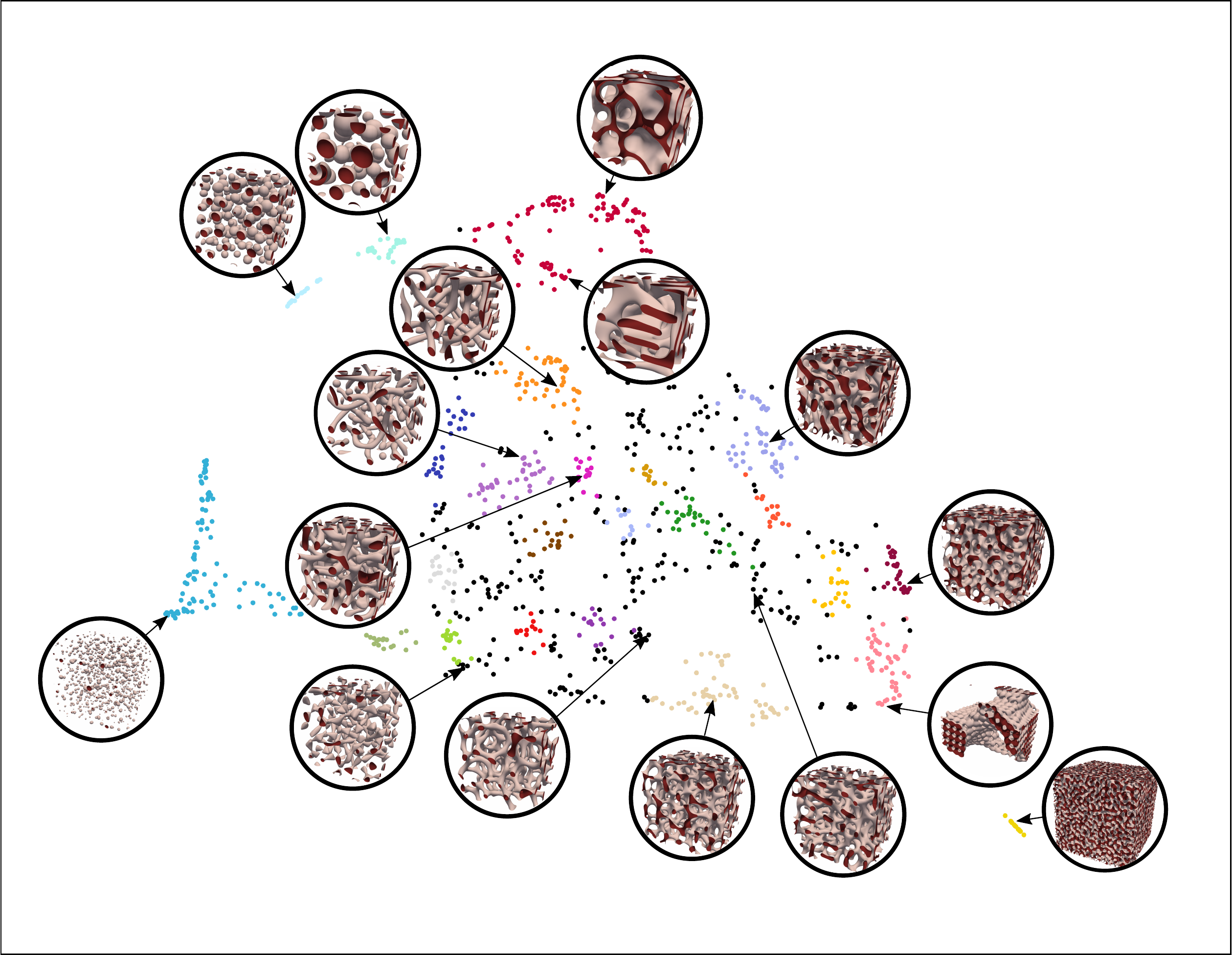}
	\caption{\small \textbf{Atlas of $1000$ shapes visualized in UMAP.} The generation parameters were chosen randomly and we applied some criteria to select mainly viable shapes. Three major families can be identified: spheres, layers, and tubules which constitute the principal type of shape textures. Let us highlight the large intra-variation in the family of tubules, that features not only smooth sponges and long tubes, but also irregular and anisotropic tubules with a multi-scale texture. This is a remarkable feature of the generation model, since all pixels of the simulation domain share identical properties with respect to the minimization problem.}
	\label{fig:atlas}
\end{figure*}

Our final experiment is designed to explore the space of possible shape textures with \verb|curvatubes| and visualize them in a 2D atlas (see Figure \ref{fig:atlas}). The generation parameters were chosen randomly. We fixed $a_{2,0} = 1$, and arbitrarily chose the other coefficients according to a uniform law in the following intervals: $a_{1,1} \in (-4,4)$, $a_{0,2} \in (1/15,15)$, $b_{1,0} \in (-200,200)$, $b_{0,1} \in (-200,200)$, $c \in (-3000,3000)$, and the mass $m_0 \in (-0.75, -0.15)$ which represented from $12.5\%$ to $42.5\%$ of relative volume occupied by the phase $\{ u \simeq 1 \}$. The initialization $A_0$ was refreshed for each simulation.

By doing this, the algorithm was pushed to its limits, as some values of coefficients chosen in this random way led to an ill-posed geometric minimization problem. Yet, even in these cases, the algorithm did not diverge to NaN values, but the function $u$ simply did not converge to a phase-field with two distinct phases, or had no zero level set, or the zero level set was not smooth and was fragmented into pieces. To reject such situations, we gauged the viability of the parameters by computing the \textit{discrepancy} $\int_\W \left|\frac{\eps}{2} \ngd^2 - \frac{W(u)}{\eps} \right| ~dx$ of the phase-field, normalized by the diffuse area, i.e., the Cahn-Hilliard energy,

\begin{equation*}
\frac{\displaystyle \int_\W \left|\frac{\eps}{2} \ngd^2 - \frac{W(u)}{\eps} \right| ~dx}{\displaystyle \int_\W \left(\frac{\eps}{2} \ngd^2 + \frac{W(u)}{\eps} \right) ~dx} \in [0,1].
\end{equation*}

The discrepancy measures how much $u$ deviates from a phase-field with $\tanh$ profile as in \eqref{eq:tanh_profile},
and is an indicator of a good behavior in numerical experiments.  
After normalization by the diffuse area, the quantity obtained varies between $0$ and $1$, with $0$ indicating a good quality in the numerical approximations.

Shapes were deemed viable if the normalized discrepancy was under the threshold $0.75$ and if $\text{max}(u) > 0.1$ and $\text{min}(u) < -0.1$ to ensure that the zero level set was defined. If the random value assigned to $(\abf,m_0)$ produced a non-viable shape, a new value was drawn uniformly until obtaining a viable shape.\newline

We thus generated $1000$ shapes meeting the criteria mentioned above, and computed the pairwise Wasserstein distance between their curvature diagrams, as described in the last paragraph of Section \ref{sec:algo_description}. 
To reduce the computation time, in each comparison we restricted the point cloud \eqref{eq:curva_diagram} to $10000$ cells randomly taken from the whole mesh.
The distance matrix was then given as input to UMAP \cite{mcinnes_umap_2020}, a manifold learning technique for dimension reduction, with the option \textit{metric = `precomputed'}. The $1000$ shapes were embedded in a 2D atlas by considering local neighborhoods of $n = 25$ points, a minimal distance $md = 0.05$ between embedded points and a spread $sp = 1$. For reproducibility, the random seed was set to $RS = 1$. 
To enhance the visualization, the embedded points were labeled with Hdbscan \cite{mcinnes_hdbscan_2017}, with a minimum cluster size $mcs = 10$ and a minimum number of samples $ms = 10$. They were colored according to their cluster number, and as black dots if Hdbscan classified them as noise. We picked some shapes from the point cloud and displayed their thumbnails, with their location specified by an arrow. Please note that the thumbnails do not exhaustively cover all the types of morphologies, but may give an idea of their disparity.

The atlas in Figure \ref{fig:atlas} shows that the shape textures are roughly distributed into three main families, spheres (top, left), layers (top), and tubules (central part) that occupy the largest region. In two marginal regions, we also identified outliers, such as highly packed tubules (bottom, right), and fragmented shapes (bottom, left). The latter suggest that the selection criteria mentioned above were not selective enough for discarding badly-converged shapes. The marginal regions concentrate most of the outliers, but we also noticed a few of them spread inside the main regions.

The transition between morphological subtypes is quite smooth when moving continuously in the atlas; however we did not examine in which way the generation parameters relate to the spatial embedding yet. The family of tubules has a large intra-variation, and features not only smooth sponges or long tubes, but also irregular, tortuous and anisotropic tubules that have a multi-scale texture, as mentioned in Experiments 2. This is a remarkable behavior of the generation model, since in regard of the minimization problem, all points have the same homogeneous properties in space.

\section{Discussion}
\label{sec:discussion}

In this final section, we discuss the strengths and limitations of the algorithm, propose a few extensions, and present the important implications of a unifying framework on applied contexts. 

\subsection{Strengths and limitations}

As seen in the simulations of Section \ref{sec:simulations}, \verb|curvatubes| leads to a wide range of membranous and tubular shapes, some of which have a multi-scale texture. The generation parameters and the curvature diagrams capture well the notion of shape texture. The algorithm is GPU-accelerated and takes advantage of automatic differentiation combined to external algorithms (Adam, L-BFGS) to descend gradient flows. Contrarily to refined numerical schemes, its aim is not to precisely solve the evolution equation, but rather to converge fast to a local minimizer with small energy. The mathematical computation of the gradient is not required either. The coefficients can be chosen in a flexible way, without letting the algorithm diverge numerically, even in mathematically ill-posed cases. The simulation results are reproducible, and seem to behave in accordance with the polynomial of curvatures in the energy especially under the form \eqref{eq:redundant}; although much work is still needed to understand mathematically how different polynomial energies are linked to different shape textures, and what are the values of coefficients that correspond to well-posed geometric problems.

Let us caution the reader that here, the model does not generate tree-like structures, for which junctions are hierarchically organized into parent and children nodes, and cycles are excluded. Thus, it cannot be applied to the respiratory system, and can only model vessels that branch and cycle a lot such as capillaries. This is because the model is devised primarily for reproducing shape texture, but not \textit{shape structure}. Some extensions of the framework to include structured constraints are proposed in the next subsection. The notion of shape texture is inspired from visual texture in images, characterized by spatially repeated elements whose conformation, such as size, color, orientation, are subject to randomness \cite{julesz_visual_1962,portilla_parametric_2000,landy_visual_2004}. Texture is hence a statistically defined property, while structure may be understood as an orthogonal component.

Furthermore, in contrast to the Helfrich biomembrane model and the FCH model of \cite{gavish_variational_2012}, which truly model some physico-chemical energy derived from microscopic interactions, we do not assume any such physical ground to the general curvature functional $\Fbf$ and the corresponding phase-field $\Feps$ that we propose. This framework is simply intended to provide a descriptive tool to analyze tubular textures, and may be used to quantify biological shapes in terms of geometry, even without any knowledge of the underlying microscopic interactions.

\subsection{Extensions}
\label{sec:discu_extensions}

\begin{figure}[h!]
	\centering
	\includegraphics[trim = 0 0 0 0, clip, width=\linewidth]{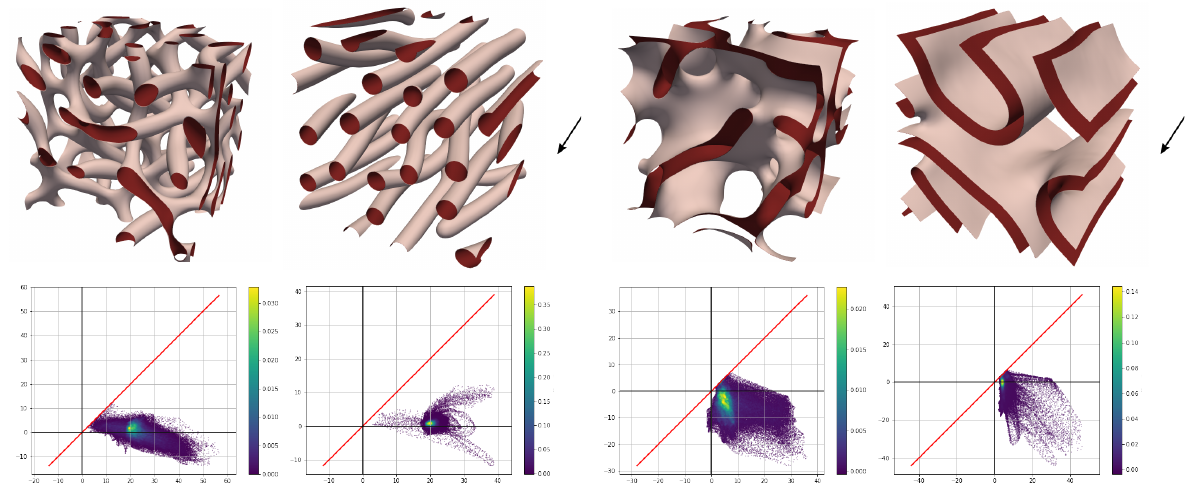} 
	\caption{\small\textbf{Including orientation in the loss aligns tubes and layers.} The tubes were generated with $\abf = (1,2,6,-40,-40,400)$, which corresponds to $\Fbf = (H-20)^2 + 5 \kb^2$, $m_0 = -0.6$, and $\mu = 0$ or $1000$ for the first and second shapes. The layers were generated using $\abf = (1,1,1,0,0,0)$, which corresponds to $\Fbf = (H^2 + \ka^2 + \kb^2)/2$, $m_0 = -0.4$, and $\mu = 0$ or $800$ for the third and fourth shapes. When $\mu \neq 0$, the direction of $\theta$ is indicated by the arrow. The curvature diagrams show that after alignment, the curvatures are more densely clustered around the dirac masses associated to perfectly cylindrical or perfectly flat shapes, namely, $\delta_{(20,0)}$ for tubes and $\delta_{(0,0)}$ for layers.
	}
	\label{fig:theta}
\end{figure}

We can include a constraint on the orientation of the normal vector $\n_\S$ to the surface $\S$, by encouraging $\n_\S$ to be orthogonal to the direction associated to a vector $\theta$ of unit norm, as in
\[ \widetilde \Fbf(\S) = \int_\S p(\ka, \kb) ~ dA + \mu  \int_\S (\theta \cdot \n_\S)^2 ~ dA.\]
This can be approximated by the phase-field energy
\[ \widetilde{\Feps} = \Feps + \mu \int_\W (\theta \cdot \n_u)^2 \, \eps \ngd^2 ~ dx 
= \Feps + \eps \mu \int_\W (\theta \cdot \grad u)^2 ~ dx, \]
where $\Feps$ is our phase-field expression constructed in Section \ref{sec:Feps_expression}.
The effect on tubes is to align their median axis along $\theta$, while inciting flat layers to be parallel to $\theta$, as in Figure \ref{fig:theta}.

Another way to give some structure to the shape is to use space-dependent generation parameters $(\abf(x),m(x))$, i.e., make them spatialized instead of constant, as in Figure \ref{fig:spatialized}. In the current version of the algorithm, since $u$ is periodic, the coefficients $\abf(x)$ but also the mass $m$ are required to be periodic.
The change of variable of Section \ref{sec:algo_description} becomes 
\[ u(x) = \Div A (x) + m(x).\]
In Figure \ref{fig:spatialized}, we took four reference generation parameters $p_1$, $p_2$, $p_3$, and $p_4$ (see Table \ref{tab:in_spatialized}), and linearly interpolated them along the horizontal axis, by taking into account the periodicity. We also repeated the first and last values (following the order $p_1$, $p_1$, $p_2$, $p_3$, $p_4$, $p_4$), and cropped the shape by dropping the first and last cubes. The spatialized parameters hence coincide with $p_1$, $p_2$, $p_3$, $p_4$ at the vertical midplanes of the four cubes delimited by the dashed lines.

\begin{table*}[h]
	\small\centering
	\begin{tabular}{|c|c|c|c|}
		\hline 
		$\abf$ & $\bbf$ & $p$ in energy $\Fbf$ & $m_0$ \\[.2em]
		\hline
		$(2,2,2,0,0,0)$ & $(1,0,0,1,0,1,0)$ & $H^2 + \kappa_1^2 + \kappa_2^2$ & $-0.4 ~(30\%)$ \\ 
		$(1, 2, 6,-40, -40, 400)$ & $(1,20,0,0,0,5,0)$ & $(H-20)^2 + 5 \kb^2 $ & $-0.6 ~(20\%)$ \\ 
		$(2, 2, 2, -75, -75, 937.5)$ & $(1,25,0,1,12.5,1,12.5)$& $(H-25)^2 + (\ka - 12.5)^2 + (\kb - 12.5)^2 $ & $-0.7 ~(15\%)$ \\ 
		$(2, 2, 11,-180,-90, 4050)$ & $(1,45,0,1,45,10,0)$& $(H-45)^2 + (\ka - 45)^2 + 10 \kb^2 $ & $-0.7 ~(15\%)$ \\ 
		\hline
	\end{tabular}
	\caption{\small\textbf{Generation parameters used in the spatialized interpolation of Figure \ref{fig:spatialized}}, associated with the central regions of the four cubes delimited by dashed lines (from left to right). The vector $\bbf$ parameterizes the polyomial expression \eqref{eq:redundant}.}
	\label{tab:in_spatialized}
\end{table*}

\subsection{Importance of a unifying theory, and future applications}

Finally, we have identified several implications that a generation model unifying tubular and membranous shapes could have in other contexts.

\begin{itemize}
	\item \textbf{Design bio-inspired shape textures}: the generation model could help in the design of bio-fabricated vascular networks for tissue regeneration \cite{sarker_3d_2018}, as well as scaffolds for bone tissue engineering \cite{fantini_method_2016} and cellular solids in architecture \cite{naboni_design_2017}, both inspired by the structure of bone trabeculae.
	
	\item \textbf{Model morphological states and trajectories}: 
	if generation parameters can be inferred from morphological states, a morphological transformation can be modeled as a trajectory in the lower-dimensional space of parameters, and then analyzed as a longitudinal trajectory \cite{durrleman_toward_2013}. In particular, the biological transformations mentioned in the Introduction would be modeled in a continuous way.
	
	\item \textbf{Provide regularization prior for tubular segmentation}: the generation model could be included as a regularizing loss in variational segmentation methods of vascular structures \cite{tyrrell_robust_2007,el-zehiry_vessel_2012,merveille_variational_2016}, to select certain tubular morphologies against others. It could also be combined with 3D reconstruction from 2D slices methods \cite{li_three-dimensional_2015,bretin_volume_2017,kim_three-dimensional_2019}.
	
	\item \textbf{Build a synthetic database of textures}: the generation algorithm could provide, at a low cost, a complete panel of synthetic textures on which to test and train vascular shape analysis methods \cite{piccinelli_framework_2009,kelch_organ-wide_2015,rust_practical_2020}, including topological analysis methods \cite{niethammer_analysis_2002,kanari_topological_2018,byrne_topological_2019}, segmentation algorithms, or microvascular blood flow simulations \cite{pozrikidis_numerical_2009,balogh_direct_2017}. It could also provide a database to research in shape and texture perception \cite{vacher_portilla-simoncelli_2020}; the way we perceive shapes is intimately linked to the way we want to quantify them.
\end{itemize}

We of course did not cover all these applications here, but intend to focus on two of the points aforementioned as future work.

The first one is to model morphological states or trajectories of biological tissues as static values or trajectories of generation parameters $(\abf(t),m(t))$, which supposes that parameters can be \textit{inferred} from shapes. This can be done naively, by visual inspection and trial-and-error; or, by first producing an atlas of reference shapes densely sampling a region with the desired morphologies, similarly to Experiment 4. Using the curvature diagram of the query shape $u_0$, the shapes closest to it in terms of the Wasserstein distance are found. We can then initialize $(\abf,m_0)$ at these values, and minimize the loss
\[ \Feps[u_0 ; \abf, m_0 ] \]
with respect to the parameters $(\abf, m_0)$ instead of the phase-field $u_0$, by using nearly the same algorithmic framework as Algorithm \ref{algo:Hminus_flow}.

The second related application is to include the curvature energy as a regularizing loss in order to segment vascular structures. The energy would then select certain tubular morphologies over others. This could be used for instance to reconstruct 3D tubular structures captured in several 2D images at different depths of a biological sample (as done in optical sectioning),
provided that there are not filaments too thin compared to the diffusion width $\eps$. The method is most effective if there is some knowledge of the shape textures that need to be segmented, so that the parameters can be appropriately tuned by inference, as previously explained.

\paragraph{Acknowledgements}
The author is thankful to Dominique Bonnet and Antoniana Batsivari for providing the images of vessels that inspired this work.
She expresses her gratitude to Anthea Monod for her guidance, to Jean Feydy for helpful insights on the numerical aspects, and to Pierre Degond for his advice. Finally she acknowledges valuable discussions with Simon Masnou, Blanche Buet and Elie Bretin.
This work was jointly funded by Imperial College London and The Francis Crick Institute through a PhD studentship.

\section*{Conflict of interest}
The author declares that she has no conflict of interest.

\section*{Appendix - Proof of the $\Gamma$-limsup}

The proof of Theorem \ref{thm:gamma_limsup} consists in showing that the sequence constructed in \cite{bellettini_approximation_2010,bellettini_approssimazione_1993} satisfies the theorem for our extended formulation $\Feps$. 

Let us first recall their construction (up to a factor $\sqrt{2}$). 
By assumption, the surface $\S = \bord E \inter \W$ is $C^2$. Let $d : \W \to \R$ be the signed distance function to $\S$, with the convention that $d$ is positive inside $E$ and negative on $\W \setminus \bar{E}$. Let $\gamma_{\eps}$ be defined on $\R$ by
\begin{equation}
\gamma_{\eps}(s) = \left\{ \begin{array}{ccc}
\tanh(\frac{s}{\sqrt{2} \eps}) & \text{if} & s \in [0,b_\eps) \\
p_\eps(s) & \text{if} & s \in [b_\eps, c_\eps) \\
+1 & \text{if} & s \in [c_\eps, +\infty) \\
- \gamma_\eps(-s) & \text{if} & s < 0
\end{array} \right.,
\end{equation}
where $b_\eps = \sqrt{2} \eps |\log \eps|$ and $c_\eps = \sqrt{2} (\eps + \eps^3 + \eps |\log \eps|)$, $a_\eps = \frac{1}{(1+\eps^2)^3}$, and $p_\eps(s) = 1 - a_\eps (s-c_\eps)^2$ is a parabolic arc connecting the graphs of $\tanh(\frac{s}{\sqrt{2} \eps})$ and the constant $+1$ on the interval $(b_\eps, c_\eps)$. 
The coefficients $a_\eps$ and $c_\eps$ ensure that $\gamma_{\eps} \in H^2(\R)$.
Now, we set 
$$u_\eps = \gamma_{\eps} \circ d \in W^{2,2}(\W).$$

We split $\W$ into three regions, $\W^{(1)}_\eps = \{0 < |d| < b_\eps \}$, 
$\W^{(2)}_\eps = \{b_\eps < |d| <  c_\eps\}$ and $\W^{(3)}_\eps = \{ c_\eps < |d| \}$ (on which $u_\eps = \pm 1$). On $\W^{(1)}_\eps$ and $\W^{(2)}_\eps$, $\grad u_\eps \neq 0$ and $\Mueps = \eps |u_\eps| B^\eps_{u_\eps}$, whereas on $\W^{(3)}_\eps$, $\grad u_\eps = 0$ and $\Mueps = 0$. Note that the region $\overline{\W^{(1)}_\eps \union \W^{(2)}_\eps} = \{ d \leq c_\eps \}$ decreases and concentrates around the surface $\S$ as $\eps \to 0$, while the complementary region $\W^{(3)}_\eps$ grows. Also, on $\W^{(2)}_\eps$, $|\grad u_\eps|$ is bounded by $p_\eps'(b_\eps) = \frac{2 \sqrt{2} \eps}{(1+\eps^2)^2}$.

As $\S$ is compact and $C^2$, there exists a tubular neighborhood $\text{Tub}(\S, d_0) = \{ d \leq d_0 \}$ of $\S$ on which $d$ is $\mathcal{C}^2$ \cite{krantz_distance_1981}, and for any point $z$ in $\text{Tub}(\S, d_0)$ the distance $d(z)$ is realized by a unique point $\pi_\S(z)$ which satisfies $z = \pi_\S(z) + d(z) N(\pi_\S(z))$. 
On the tubular neighborhood, the signed distance satisfies the eikonal equation $|\grad d| = 1$, implying that $\Hess d ~ \grad d = 0$. The symmetric matrix $-\text{Hess} ~ d \,(x)$\footnote{The minus sign correspond to the convention that if $\bord E$ is a sphere, the eigenvalues should be positive.} has two eigenvalues $\lambda_1(x) \geq \lambda_2(x)$ corresponding to the principal directions of the associated level set, and a third eigenvalue $0$ in the direction $\pm \grad d$. The eigenvalues $\lambda_1$ and $\lambda_2$ are continuous on $\text{Tub}(\S, d_0)$.
\newline 

To show \eqref{eq:limsupa}, as $|u_\eps - (2 \chi_E - 1)| \leq 1$, 
\begin{align*}
\int_{\W} |u_\eps - (2 \chi_E - 1)| \, dx &= \int_{ \{ |d| < c_\eps \} } |u_\eps - (2 \chi_E - 1)| \, dx \\
&\leq \mathcal{L}^3( \{ |d| < c_\eps \}) \to 0
\end{align*}
so $\lim\limits_{\eps \to 0^+} u_\eps = 2 \chi_E - 1 \text{ in } L^1(\W)$ and it remains to prove \eqref{eq:limsupb} and \eqref{eq:limsupc}.\newline

Let $\phi \in C^0_c(\W)$ be a continuous function with compact support. We need to show that $\int_\W \eps |\grad u_\eps|^2 \, \phi ~dx $ converges to $\sigma \int_{\S} \phi ~d\hauss^2$ as $\eps$ is sent to zero. First, we work only on $\W^{(1)}_\eps$, as
\[ \int_\W \eps |\grad u_\eps|^2 \, \phi ~dx = \int_{\W^{(1)}_\eps} \eps |\grad u_\eps|^2 \, \phi ~dx + 
\int_{\W^{(2)}_\eps} \eps |\grad u_\eps|^2 \, \phi ~dx\]
and the second integral tends to zero since the integrand is bounded and $\mathcal{L}^3(\W^{(2)}_\eps) \to 0$.
By the co-area formula,
\begin{align*}
\int_{\W^{(1)}_\eps} \eps |\grad u_\eps|^2 \, \phi ~dx &= \int_{-1}^{u_\eps(b_\eps)} \left( \int_{ \{u_\eps = t\} } \eps |\grad u_\eps| \, \phi ~d\hauss^2 \right) \,dt \\
&= \int_{-1}^{u_\eps(b_\eps)} \sqrt{2 W(t)}  \left( \int_{ \{u_\eps = t\} } \phi ~d\hauss^2 \right) \,dt,
\end{align*}
where we use $|\grad u_\eps| = \frac{(1-u_\eps^2)}{\sqrt{2} \eps} = \frac{\sqrt{2 W(t)}}{\eps}$ on $\W^{(1)}_\eps$.
Therefore,
\begin{align} \label{eq:level_sets}
&\left| \int_{\W^{(1)}_\eps} \eps |\grad u_\eps|^2 \, \phi ~dx - \sigma \int_{\S} \phi ~d\hauss^2 \right| \\
&\leq \int_{-1}^{u_\eps(b_\eps)} \sqrt{2 W(t)} \left| \int_{ \{ d = \gamma_\eps^{-1}(t) \} } \phi ~d\hauss^2 - \int_{ \{ d = 0 \} } \phi ~d\hauss^2 \right| \,dt\\
&+ \int_{ \S } \phi ~d\hauss^2 ~ \int_{u_\eps(b_\eps)}^1  \sqrt{2 W(t)} \,dt,
\end{align}
where $\gamma_\eps^{-1} : (-1,1) \to ( -c_\eps, c_\eps)$ denotes the inverse of the restriction of $\gamma_\eps$ to $( -c_\eps, c_\eps)$. Using $u_\eps(b_\eps) = \frac{1 - \eps^2}{1 + \eps^2} \to 1$, the last term goes to zero.

The convergence to zero of the bound is proved if for any $\delta > 0$, we can find $\eps_0$ small enough such that $\forall \eps < \eps_0$, $\forall s \in (-c_\eps, c_\eps)$,
\begin{equation} \label{eq:ineq}
\left| \int_{ \{d = s \} } \phi ~d\hauss^2 - \int_{ \S } \phi ~d\hauss^2 \right| \leq \delta.
\end{equation}
This is true because, if $p \in \S$, there is a $C^1$-diffeomorphism $\psi$ mapping $U \times (-d_0, d_0)$ onto $\text{Tub}(V, d_0)$, where $U \subset \R^2$ is open and $V \subset \S$ is an open neighborhood of $p$.
For $s \in (-d_0,d_0)$, by the change of variables formula,
\begin{align*}
&\int_{ \{d = s \} \inter \text{Tub}(V, d_0) } \phi ~d\hauss^2 \\ 
&= \int_U \phi(\psi(x,y,s)) \sqrt{|\det(\text{Jac}_\psi^T \, \text{Jac}_\psi)|}(x,y,s) \, dx dy 
\end{align*}
where $\text{Jac}(\psi) = (\frac{\partial \psi}{\partial x}, \frac{\partial \psi}{\partial y}) \in \R^{3 \times 2}$,
and this integral converges to $\int_{ V } \phi ~d\hauss^2 $ when $s \to 0$. By compactness, we can consider a finite number of such neighborhoods $V$ and conclude that the limit \eqref{eq:limsupb} holds.\newline

Now, we prove the convergence of the energies. It can be shown that on $\W^{(1)}_\eps$, we have 
\begin{align*}
\Bueps &= -\Hess d \\
\Nor{\Bueps}^2 &= \lambda_1^2 + \lambda_2^2 \\
\Tra{\Bueps} &= -\Lap d = \lambda_1 + \lambda_2
\end{align*}
and on $\W^{(2)}_\eps$
\begin{align*}
\Bueps &= -\Hess d + f_\eps ~\n_d \otimes \n_d \\
\Nor{\Bueps}^2 &= \lambda_1^2 + \lambda_2^2 + f_\eps^2 \\
\Tra{\Bueps} &= -\Lap d + f_\eps = \lambda_1 + \lambda_2 + f_\eps
\end{align*}
where we define the auxiliary function
\begin{equation*}
f_\eps = \frac{1}{p_\eps'(d)} \left( p_\eps''(d) - \frac{W'(p_\eps(d))}{\eps^2}  \right).
\end{equation*}
such that $f_\eps |\grad u_\eps| \leq \frac{2}{(1 + \eps^2)^3} + \frac{4}{1+\eps^2}$ remains bounded on $\W^{(2)}_\eps$.\newline

It is sufficient to show that the limit holds independently for each term of $F$. We prove it for the term $\ka^2$ associated to $\abf = (1,0,0,0,0,0)$. The phase-field writes
\begin{align*}
\Feps = \frac{1}{2 \eps} \int_\W \left( \Nor{\Mueps}^2 + \Tra{\Mueps}  \sqrt{(2 \Nor{\Meps}^2 - (\Tra{\Meps})^2 )^+} \right) ~dx.
\end{align*}
Using the relationships between $\Meps$, $\Bueps$ and $\Hess d$, on $\W^{(1)}_\eps$ we have $2 \Nor{\Meps}^2 - (\Tra{\Meps})^2 = (\lambda_1 - \lambda_2)^2 \geq 0$ and
\[ \Nor{\Mueps}^2 + \Tra{\Mueps}  \sqrt{(2 \Nor{\Meps}^2 - (\Tra{\Meps})^2 )^+} 
= 2 \eps^2 |\grad u_\eps|^2 \lambda_1^2, \]
while on $\W^{(2)}_\eps$, we have $\xi := 2 \Nor{\Meps}^2 - (\Tra{\Meps})^2 = \lambda_1^2 + \lambda_2^2 + f_\eps^2 - 2 \lambda_1 \lambda_2 - 2 \lambda_1 f_\eps - 2 \lambda_2 f_\eps$. As $\xi^+ \leq |\xi| \leq 3 (\lambda_1^2 + \lambda_2^2 + f_\eps^2) \leq 3 (|\lambda_1| + |\lambda_2| + |f_\eps|)^2$, we get the bound
\begin{equation*}
| \Tra{\Mueps}  \sqrt{(2 \Nor{\Meps}^2 - (\Tra{\Meps})^2 )^+} | \leq \sqrt{3} (|\lambda_1| + |\lambda_2| + |f_\eps|)^2.
\end{equation*}

$\Feps$ can thus be decomposed into $\Feps = I_\eps + J_\eps$,
where 
\begin{align*}
I_\eps &= \int_{\W^{(1)}_\eps \union \W^{(2)}_\eps} \eps |\grad u_\eps|^2 \lambda_1^2(x) ~dx,\qquad \text{ and } \\
|J_\eps| &\leq \int_{\W^{(2)}_\eps} \eps |\grad u_\eps|^2 \left(\lambda_2^2 + f_\eps^2 + \sqrt{3} (|\lambda_1| + |\lambda_2| + |f_\eps|)^2  \right) ~dx.
\end{align*}
On $\W^{(2)}_\eps$, the functions $|\grad u_\eps|$, $f_\eps |\grad u_\eps|$, $|\lambda_1|$ and $|\lambda_2|$ are all bounded.
Therefore, as $\mathcal{L}^3(\W^{(2)}_\eps) \to 0$, we get $J_\eps \to 0$.

Finally, since $|\grad u_\eps| = 0$ on $\W^{(3)}_\eps$,
\[ I_\eps = \int_\W \eps |\grad u_\eps|^2 \widetilde{\lambda_1}^2 ~dx,\]
where $\widetilde{\lambda_1} \in \mathcal{C}^0_c(\W)$ continuously extends $\lambda_1$ beyond $\text{Tub}(\S, d_0)$. Due to the convergence of Radon measures \eqref{eq:limsupb}, this integral converges towards 
\[ \lim\limits_{\eps \to 0} I_\eps =  \sigma \int_{\S} \ka^2 ~d\hauss^2.\]
We can similarly prove the convergence for the other monomials, and therefore conclude that Theorem \ref{thm:gamma_limsup} holds.

{\small
\bibliographystyle{spmpsci}      
\bibliography{All_references}   
}

\end{document}